\documentclass[12pt]{article}
\input epsf.tex
\usepackage{amssymb}

\newcommand{\bmat}{\left(\begin{array}}
\newcommand{\emat}{\end{array}\right)}

\def\harr#1#2{\smash{\mathop{\hbox to .3in{\rightarrowfill}}
 \limits^{\scriptstyle#1}_{\scriptstyle#2}}}

\def\yzero{\smash{\hbox{$y\kern-4pt\raise1pt\hbox{${}^\circ$}$}}}
\def\ov{\overline}
\def\s2{\frac{1}{\sqrt2}}

\def\beq{\begin{equation}}
\def\eeq{\end{equation}}
\def\beqa{\begin{eqnarray}}
\def\eeqa{\end{eqnarray}}

\def\Dsl{\,\raise.15ex\hbox{/}\mkern-13.5mu D} 

\def\IR{\mathbb{R}}
\def\IC\mathbb{C}
\def\IZ{\mathbb{Z}}
\def\ItZ{\widetilde{\mathbb{Z}}}
\def\IZdos{{\bf Z}_2}

\def\IS{\mathbb{S}}

\def\IRP{\mathbb{R}{\rm P}}

\def\Oplus{{O3^+}}
\def\Ominus{{O3^-}}
\def\Otplus{{\widetilde{O3}^+}}
\def\Otminus{{\widetilde{O3}^-}}
\def\Dseven{{\ov{D7}}}

\topmargin
0cm
\textwidth
15.5cm
\textheight
23cm
\oddsidemargin
0.7cm
\evensidemargin
1.2cm

\begin{document}

\makeatletter
\@addtoreset{equation}{section}
\makeatother
\renewcommand{\theequation}{\thesection.\arabic{equation}}
\pagestyle{empty}
\rightline{CTP-MIT/3434}

\rightline{\tt hep-th/0311028}
\vspace{.5cm}
\begin{center}
\Large{\bf Instantonic Branes, Atiyah-Hirzebruch Spectral Sequence, and $SL(2,\IZ)$ duality of ${\cal N}=4$ SYM}\\
\vspace{1cm}

\large
Oscar Loaiza-Brito\footnote{e-mail address:{\tt oloaiza@lns.mit.edu}} \\[2mm]

{\em Center for Theoretical Physics}\\
{\em Massachusetts Institute of Technology}\\ 
{\em Cambridge, Massachusetts 02139, U.S.A.}\\[4mm]

\vspace*{2cm}
\small{\bf Abstract} \\[7mm]
\end{center}

\begin{center} 
\begin{minipage}[h]{14.0cm} {We study the equivalence among orientifold three-planes in the context of the Atiyah-Hirzebruch Spectral 
Sequence. This equivalence refers to the fact that two different cohomology classes in the same cohomology group, which classify the orientifolds, 
are lifted to the same trivial class in K-theory. The physical interpretation of this mathematical algorithm is given by the role of D-brane 
instantons. By following some recent ideas, we extend the sequence to include a classification of NS-NS fluxes. We find that such 
equivalences, in the low energy limit of the dynamics on the worldvolume of type IIB $D3$-branes on top of the orientifolds, are 
interpreted as the $SL(2,\IZ)$ duality in four dimensional ${\cal N}=4$ SYM theories. }
\end{minipage} 
\end{center}

\bigskip

\bigskip
  

\leftline{November 2003}

\newpage
\setcounter{page}{1}
\pagestyle{plain}
\renewcommand{\thefootnote}{\arabic{footnote}}
\setcounter{footnote}{0}

\section{Introduction}
There are a variety of orientifold planes for each dimensionality $p$ related to the fact that it is possible to turn on discrete RR and
NS-NS fluxes \cite{WB, HO, HT, BGS}. These fluxes, a NS-NS three form $H_{NS}$ and a R-R form $G_{6-p}$, are classified by integral  cohomologies
$H^3(\IRP^{8-p})$ and $H^{6-p}(\IRP^{8-p})$ respectively (both with a discrete torsion value $\IZdos$). However, such a classification does
not provide neither a satisfactory explanation of the fractional RR charge carried by orientifold planes for $p\leq 4$ nor the relative
charge between a pair of them with the same dimensionality $p$ (actually the one with $[H_{NS}]=[G_{6-p}]=[0]$ and the one with $[H_{NS}]=[0]$ and
$[G_{6-p}]=[1]$).

In reference \cite{BGS}, Bergman, Gimon and Sugimoto provide a K-theory classification of orientifolds which prove to resolve the latter problem
and also give us a more accurate picture of the non-perturbative nature of them. One of the most interesting phenomena arising from
this classification is the fact that there are not four types of, for instance, orientifold three-planes as cohomology suggests, but only
three. This means that two $O3$-planes, which seemed to be different objects in a cohomological classification, turn out to be the
same object in K-theory (the same situation happens for orientifolds with dimensionality $p\leq 3$). Mathematically, this is a consequence of
the comparison between cohomology and K-theory in the classification of RR fluxes, given by the Atiyah-Hirzebruch Spectral Sequence (AHSS). 
 
On the other hand, by asking for suitable consistent conditions to wrap a $D$-brane on non-trivial homology cycles of spacetime, Maldacena,
Moore and Seiberg (MMS) \cite{M}, showed that it is possible that a Freed-Witten anomaly-free system \cite{FW} (see also \cite{Wk, DMW}) consisting 
in a D-brane wrapping a
non-trivial cycle $\cal W$ can be nevertheless unstable to decay into vacuum if there is an anomalous system, formed by a $D$-brane
wrapping a non-trivial cycle $\Sigma$, such that $\cal W$ is contained in $\Sigma$ and $codim\;{\cal W}|_\Sigma =3$. This is indeed a 
physical
picture of the AHSS classifying RR fluxes(reviewed herein).

In this paper we apply the MMS proposal in the presence of orientifold three-planes by means of which we give a physical picture of the
equivalence of $O3$-planes given previously in \cite{BGS}. Also we extend this proposal to include a classification of NS-NS fluxes at the 
level
of the comparison between cohomology and K-theory, and we find that there are some other equivalences between different $O3$-planes. However, 
since there are different gauge field theories related to the four types of orientifold planes (classified by cohomology), there must be a 
relation between the former ones as a consequence of the orientifold equivalences in K-theory.  
Interesting
enough, we find that such equivalences (two different non-trivial cohomology classes belonging to the same cohomology group, are both 
lifted to the trivial class in K-theory) are interpreted in the low energy limit of the field content on $D3$-branes on top of the 
orientifold
three-planes, as the well-known $SL(2,\IZ)$ duality in four dimensional ${\cal N}=4$ Super Yang-Mills theories. It is interesting to know 
that similar results were found in ref. \cite{gerbes} in a classification of three-orientifolds by flat gerbs\footnote{I thank Arjan
Keurentjes for pointing me out these references.}.

Our paper is organized as follows: in section 2 we give an overview of four-dimensional SYM theories from the point of view of orientifolds 
and 
$D3$-branes, the construction of the four types of orientifold three-planes by (co)homology and the $SL(2,\IZ)$ duality present in these
systems. In section 3 we start by summarizing the MMS picture of the AHSS. The AHSS is reviewed in Appendix A. Later on, we recall the equivalence between $\Oplus$ and $\Otplus$ in
the AHSS context and then we apply the MMS picture to this equivalence. We find that, by means of it, the ${\cal T}\in SL(2,\IZ)$ duality is
generated in the (massless) field content on $D3$-branes. 

In section 4 we study the generalization of the AHSS in order to include NS-NS fluxes and the corresponding MMS picture. In this way, we are
able to generate the full $SL(2,\IZ)$ duality on the massless fields on the worldvolume of $D3$-branes. 
In section 5 we study the case for an $\Ominus$ (gauge theories with $G=SO(2N)$ and the case without orientifolds ($G=U(N)$). 
Finally, we give our conclusions.

\section{${\cal N}=4$ SYM, D3-branes and Orientifolds}
It is well-known (see for instance \cite{GK, Be}) that four dimensional ${\cal N}=4$ SYM can be described by the low energy limit of the dynamics
on the worldvolume of $N$ parallel $D3$-branes in type IIB string theory. The gauge group related to this theory is $U(N)$, while symplectic
and orthogonal groups are related to $D3$-branes which are parallel to orientifold three-planes. In general, there are four different
types of orientifold three-planes which can be studied in this context. They are labeled as $\Ominus, \Oplus, \Otminus$ and $\Otplus$.
The
upper signs $\pm$ in $\Oplus$ and $\Ominus$ stand for the positive or negative RR charge they carry. The RR charge for these two $O3$-planes 
is $2^{p-5}$ in units of
D-branes. They are the T-dual versions of nine-dimensional orientifold planes which in turn are part of the type I string theory
($O9^-$) and the $USp(32)$ string theory \cite{USP32} ($O9^+$), respectively, where the latter one is a non-supersymmetric theory. Recall that 
string
theories with ${\cal N}=1$ are obtained by gauging away half of the fields present in type IIB string theory. 

The presence of an orientifold plane establishes an action over the fields in the theory which do not depend whether the orientifold is 
positive or
negative-charged. However, only fields which are even under the $O9^-$-plane action will survive the projection and they will be present
in the spectrum of type I string theory. The opposite holds in the case of the $USp(N)$ string theory, where the fields which were projected
out in the former case, will be present in this theory. One example of this, is the NS-NS two-form field $B_{NS}$, which is odd under the
action of the nine-orientifold planes, and then it is present in the field content in type $USp(32)$ theory but not in type I. After taking
T-duality over these string theories, $B_{NS}$ keeps being odd under the action of lower dimensional orientifold planes. This situation is
afterwards, not true for its magnetic  dual $\widetilde{B_6}=*B_{NS}$ for it depends on the orientifold-plane dimensionality. Such a field is odd (even) 
if the
orientifold dimension is odd (even). In the presence of an orientifold three-plane, both fields, $B$ and $B_6$ are odd.

A slightly more elaborated condition holds also for RR fields (see \cite{HO, BGS}). The action of an orientifold three-plane over RR fields is
given by
\beqa
\begin{array}{rl}
C_q \rightarrow C_q & \hbox{if } q\,=\,0\,mod\,4\\
C_q \rightarrow -C_q & \hbox{if } q\,=\,2\,mod\,4\:.
\label{rrfields}
\end{array}
\eeqa
The forms which are odd under an orientifold action are called {\it twisted} forms. Hence, the NS-NS $B$-field, its magnetic 
dual,
and the RR fields $C_6$ and $C_2$ , are all twisted forms. At this point, it is
plausible to classify their field strengths by the cohomology of the transverse space to the orientifold. 
This is the projective compact space $\IRP^5=\IS^5/\IZdos$.

So, twisted forms are classified by twisted cohomologies of $\IRP^5$. In particular the 
field strengths $G_3=dC_2$ and $H_{NS}=dB_{NS}$ are both classified by the same cohomology group, $H^3(\IRP^5;\ItZ)=\IZdos$. The physical meaning of
these {\it torsion values} is gathered via the so-called ``brane realization of discrete torsion" (see \cite{HO, HT, BGS, braneboxes}). Consider for instance a
$NS5$-brane intersecting an $O3^-$-plane. The NS-NS charge of the five-brane is a half of the same brane non-intersecting the orientifold. 
When we
integrate the field strength $H_{NS}$ over a three cycle, one gets a value of $1/2\, mod\, 1$. This is essentially the discrete torsion value
given by cohomology. However, such a value establishes a phase in the $\IRP^2$ amplitude of the orientifold plane (the M\"obius strip) which
changes the conditions to project out the type IIB fields (in order to construct a theory with 16 generators of supersymmetry over the
worldvolume of a $D3$-brane on top of the orientifold). The result is that we have a different orientifold plane which is denoted as
$O3^+$. The physical meaning is that the RR charge the orientifold was
carrying before the intersection with the $NS5$-brane reverses its sign after the intersection, i.e., the discrete torsion value of
$H_{NS}$ stands for the interchange between $\Ominus$ and $\Oplus$ -planes.

A similar situation is obtained when a $D5$-brane intersects an $O3$-plane although we can only describe it in a non-perturbative way. 
In this case the NS-NS field strength also
acquires a discrete torsion value after being integrated over a three-cycle. This RR discrete torsion is related to a different orientifold
plane denoted as $\widetilde{O3}$. Hence, there are, besides the two $O3^\pm$-planes, other two planes denoted by $\widetilde{O3^\pm}$.

So for instance, an $\Otplus$-plane is related to a non-trivial value of $H_{NS}$ and $G_3$. In general, we classify the orientifolds
according to the values of cohomology classes $([G_3],[H_{NS}])=(\theta_R, \theta_{NS})$, with $[H_{NS}]$ and $[G_3]$ both belonging to the
cohomology group $H^3(\IRP^5,\ItZ)=\IZdos$. Some of their properties are listed in table \ref{o3planes} (notice that in this case, $\Otminus$
has a positive RR charge).

\begin{table}
\begin{center}
\caption{Orientifold three-planes and some of their properties.} 
\label{o3planes}
\begin{tabular}{||c|c|c|c||}\hline\hline
$(\theta_R, \theta_{NS})$&             &Gauge Group&  Charge\\\hline\hline
$(0,0)$                  & $\Ominus$   &$SO(2N)$   & $-1/4$\\
$(1,0)$                  & $\Otminus$  &$SO(2N+1)$ & $+1/4$\\
$(0,1)$                  & $\Oplus$    &$USp(2N)$  & $+1/4$\\
$(1,1)$                  & $\Otplus$   &$USp(2N)$  & $+1/4$\\\hline\hline
\end{tabular}
\end{center}
\end{table}

A $U(N)$ gauge-theory is gathered by considering the gauge field content of $N$ D3-branes (i.e., without the inclusion of
orientifolds). This field theory is particularly interesting since it is selfdual under $SL(2,\IZ)$-transformations. One of the corresponding
generators of this symmetry group, inverts the gauge coupling constant $\tau \rightarrow -\frac{1}{\tau}$ giving us the possibility to study
the non-perturbative sector from a perturbative perspective. This behavior is also present in the four-dimensional SYM-theory with an orthogonal
gauge group $SO(2N)$ which is selfdual under the symmetry group $SL(2,\IZ)$ and is given by $2N$ D3-branes on top of an $O3^-$-plane. On the
other hand, the gauge field content obtained by putting D3-branes on top of the other three-orientifold planes are not selfdual for
each single case, but they are related to each other by the $SL(2,\IZ)$-symmetry group. 
Basically, in this section we review some important aspects of this relationship between different gauge SYM-theories (from a
brane-perspective) and more important for the rest of this note, we explain the mechanism we use to build up the four different types of
orientifold three-planes.

\subsection{Orientifold three-planes and homology}
As was said, there are four types of orientifold three planes. An alternative (mathematically equivalent) way to realize the existence of
these orientifold planes, is by studying homologically non-trivial cycles of $\IRP^5$ where we can wrap type IIB $D$-branes\footnote{We
shall see in next section that physically, we require more conditions to wrap a $D$-brane on a non-trivial cycle. In this sense, this is the
first approximation of a 'more physical' model.} (see \cite{HIS, yo, yo2}). We are
interested in obtain three dimensional objects by wrapping different types of $D$-branes on non-trivial cycles. However, there are certain
restrictions we must keep in mind. The first one is that those RR fields which coupled to type IIB $D$-branes are affected by the orientifold
three-plane projection, giving them the label of twisted or untwisted forms. So, $D$-branes can wrap, for instance a twisted (untwisted) 
homology cycle
if and only if the RR field associated to this brane is also a twisted (untwisted) form. 
In our case, as can be read off from equation (\ref{rrfields}), $D7$, $D3$ and $D(-1)$ branes wrap untwisted cycles, while $D5$ and $D1$ wrap
twisted ones. So, in order to get a three-dimensional object, after the `wrapping process', we require to wrap a $D7$-brane on a 
untwisted 4-cycle, or a $D3$-brane on a untwisted 0-cycle. Also, we can wrap a $D5$-brane on a twisted 2-cycle.

However, (and this is the second restriction) only the twisted two-cycle and the untwisted zero one are non trivial. Let us study the former cycle. Since
$H_2(\IRP^5;\ItZ)=\IZdos$ we can say that by wrapping a $D5$-brane on
$\IRP^2\subset\IRP^5$ we get a three dimensional object with discrete torsion charge $\IZdos$. Applying Poincare duality to this homology
group, we get the well-know cohomology group $H^3(\IRP^5;\IZdos)$ which classifies $O3$ and $\widetilde{O3}$ -planes. Hence, afterwards we
are classifying $O3$-planes by wrapping branes on non-trivial homology cycles of the transverse space to $O3^-$. Notice also, that we can do
the same for a $NS5$-brane, since $H_6$ is a twisted form. By wrapping a $NS5$-brane on $\IRP^2$ we get a three dimensional object with a
torsion value $\IZdos$. The trivial class in this discrete group stands for $\Ominus$ while the non-trivial one stands for $\Oplus$.

Let us turn our attention to the untwisted zero-cycle. First of all notice that $H_0(\IRP^5;\IZ)=\IZ$, so all objects we got by 
wrapping $D$-branes on a zero-cycle, would have an integer charge. In fact, by wrapping a $D3$-brane on this zero-cycle,
we get the charge of $D3$-branes on top of an $O3$-plane, a result that was expected. Applying Poincare duality, we get the cohomology group
$H^5(\IRP^5;\IZ)$ which in turn, classifies the field strength $G_5=dC_4$, with $C_4$ the RR field which couples to $D3$.

\subsection{Four dimensional $SL(2,\IZ)$ duality and Orientifold Three-planes}
The field content obtained by the low energy limit of the dynamics on $D3$-branes, is actually a ${\cal N}=4$ SYM theory with gauge group
$U(N)$. As it is well-known (see for instance \cite{monopoles}), this theory depends on a complex parameter $\tau$ which involves the SYM gauge coupling $g_{SYM}$ and the
$\theta$-term, introduced by Witten. This coupling constant is 
\beqa
\tau = \frac{\theta}{2\pi} + \frac{1}{g^2_{SYM}}
\eeqa
By the periodicity of $\theta$, a monopole acquires an electric charge when this term is considered. Also the theory is symmetric under
transformations $\tau \rightarrow \tau +1$.  In addition, the theory has other symmetry (Olive's strong-weak duality) which consists in
inverting the complex coupling $\tau \rightarrow -1/\tau$. Both together generate the full symmetry group $SL(2,\IZ)$ under which, 
the SYM theory
with 16 generators of supersymmetry, is invariant. The generic action of this group on $\tau$ is
\beqa
\tau \rightarrow \frac{p\tau +r}{q\tau +s}\;,
\eeqa
where $ps-qr=1$. Electric and magnetic charges transform under $SL(2,\IZ)$ as
\beqa
\left(
\begin{array}{c}
e'\\
m'
\end{array}
\right) \;=\;
\left(
\begin{array}{cc}
p&r\\
q&s\\
\end{array}
\right)
\left(
\begin{array}{c}
e\\
m
\end{array}
\right)\;.
\eeqa
Since the same invariance under $SL(2,\IZ)$ is present in type IIB theory, actually the field strengths $G_3$ and $H_{NS}$ satisfy the same
transformation rule under the duality group. When we have orientifolds in the theory, as we saw, the values of the field strengths $G_3$ 
and $H_{NS}$
are defined up to integers $mod\;2$. Hence, if we transform these fields under a $SL(2,\IZ)$ transformation we get (see \cite{EGKT, U}), 
\beqa
\left(
\begin{array}{c}
G_3\\
H_{NS}\\
\end{array}
\right)\rightarrow 
\left(
\begin{array}{c}
pG_3+rH_{NS}\\
qG_3+sH_{NS}
\end{array}
\right)\;=\;
\left(
\begin{array}{c}
G_3'\\
H_{NS}'
\end{array}
\right)\;,
\eeqa
where we have used the fact that
both fields are in the same cohomology group, which allow us to express them in terms of the new fields $G_3'$ and $H_{NS}'$.

\begin{figure}
\begin{center}
\centering
\epsfysize=7cm
\leavevmode
\epsfbox{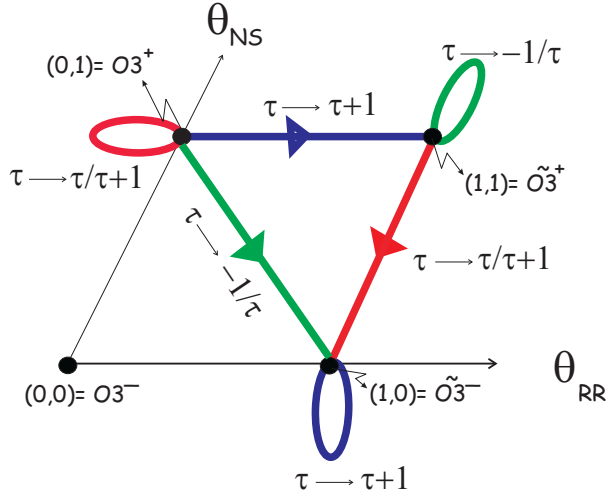}
\end{center}  
\caption[]{\small $SL(2,\IZ)$ transformations on orientifold three planes. The loops represent the selfduality of orientifolds under the
correspondent transformation.}
\label{03dualities}   
\end{figure}

Since each type of orientifold three-plane is defined by these pair of fields, we get that $\Ominus$, $\Oplus$ and $\Otplus$ transform to each
other by a $SL(2,\IZ)$ transformation. In particular we see that for $p=r=s=1$ and $q=0$ ($\tau \rightarrow \tau+1$), $\Oplus$ is
transformed into $\Otplus$. In the same token, under S-duality, $\tau \rightarrow -1/\tau$, $\Oplus$ is transformed into 
$\Otminus$. Notice that $\Ominus$ is selfdual under any $SL(2,\IZ)$-transformation. The above relationships are depicted in 
figure \ref{03dualities}.

Finally, the difference in the field theory on a $D3$-brane parallel to an $\Oplus$ or an $\Otplus$ (${\cal N}=4$ SYM with
gauge group $USp(2N)$ in both cases) is given by the
spectrum of dyons and monopoles (see \cite{HO}). Using the above $SL(2,\IZ)$ relations it is shown that the root latices, where monopoles 
and dyons
belong, are also interchanged under such transformations. Let $C$ be the root lattice of the gauge group $USp(N)$, hence, electric
charges in both theories belong to $C$. Let $B$ be the root lattice for the gauge group $SO(2N+1)$. Under S-duality, electric charges from the
theory with an $\Oplus$ -plane are interchanged with monopoles in the theory with an $\Otminus$-plane. So, monopoles related to the
$\Oplus$-plane belong to the root lattice $B$. Now under $\tau \rightarrow\tau +1$, we know that monopoles acquire electric charge, so they
become dyons. Then, dyons related to an $\Otplus$-plane are interchanged with monopoles on an $\Oplus$-plane. Hence, the difference between a
field theory in the presence of an $\Oplus$ and an $\Otplus$ lies in the root lattices and the monopoles and dyons which belong to them. 
The results are shown
in table \ref{o3dyons}, where $D$ stands for the root lattice of the gauge group $SO(2N)$.

\begin{table}
\begin{center}
\caption{Spectrum of $1/2$ BPS states in the presence of $O3$-planes.} 
\label{o3dyons}
\begin{tabular}{||c|c|c|c|c||}\hline\hline
Orientifold plane        & $\Ominus$   &$\Oplus$   & $\Otplus$&$\Otminus$\\\hline\hline
Gauge group              & $SO(2N)$    &$USp(2N)$  & $USp(2N)$&$SO(2N+1)$\\
Electric charges         & $D$         &$C$        & $C$      &$B$\\
Monopoles                & $D$         &$B$        & $C$      &$C$\\
Dyons                    & $D$         &$C$        & $B$      &$C$\\\hline\hline
\end{tabular}
\end{center}
\end{table}

\section{The equivalence between $\Oplus$ and $\Otplus$. An alternative picture for ${\cal T}\in SL(2,\IZ)$-duality}
In \cite{BGS} was shown that $\Oplus$ and $\Otplus$ are equivalent in a K-theory classification. This equivalence refers to
the fact that the cohomology classes which classify both $O$-planes (actually the trivial and non-trivial classes of 
$[G_3]\in H^3(\IRP^5;\ItZ)$), 
are lifted to the same trivial class in K-theory. The reason of that lies in the context of the Atiyah-Hirzebruch Spectral Sequence (AHSS)
(see appendix B).
This sequence was physically interpreted by Maldacena, Moore and Seiberg (MMS) in \cite{M} where equivalences in K-theory were interpreted as a
consequence of the presence of instantonic branes. In this section we apply the MMS picture to the
equivalence of the above two orientifold three-planes. Importantly enough, the resulting picture give us an alternative way to understand the
duality in four dimensional ${\cal N}=4$ SYM theories under the transformation $\tau \rightarrow \tau -1$ via the presence of an instantonic
seven-brane. Our conclusion is that a K-theory equivalence between these two objects implies the presence of the $\cal T$-duality in the 
field content
gathered from D3-branes in the presence of such orientifold planes. In next section we shall see that we are able to extend this picture 
to the full $SL(2;\IZ)$ symmetry,
including of course, the weak-strong duality or S-duality. 

We start by giving a review of the MMS picture and later we give all the details of our construction of the equivalence between $\Oplus$
and $\Otplus$ in the context of AHSS and the MMS picture as well.

\subsection{The MMS picture}
The AHSS is an algebraic algorithm that essentially relates cohomology to K-theory through its graded complex $GrK(X)$(see appendix B). 
In Ref. \cite{M} (see also \cite{Mo}) MMS gave a physical interpretation to this algorithm by studying the required conditions to wrap 
a $D$-brane on a submanifold of the spacetime $X$. Here we give a review of this proposal.

Let ${\cal W}_p$ be a $p$ cycle of $X$ ( a $p$-dimensional submanifold) where a D-brane is wrapped. This cycle defines a class 
$[{\cal W}_p] \in
H_p(X;\IZ)$. Applying Poincare duality, $H_p(X;\IZ) \cong H^{9-p}(X;\IZ)$ ($dim\: X =9$). Hence $[{\cal W}_p]$ is related to a $(9-p)$-form
$\omega_{9-p}\in H^{9-p}$. The first approximation to the graded complex in the AHSS is given by $H^{9-p}(X;\IZ)$.

Now, we proceed to the second level of approximation, given by the cohomology of the differential $d_3$, where $d_3: H^p \rightarrow
H^{p+3}$. The form $\omega_{9-p}$ is closed under $d_3$ if
$d_3 (\omega_{9-p})=0$. On the other hand, $\omega_{9-p}$ is exact if $\omega_{9-p}=d_3(\sigma_{6-p})$, where $\sigma_{6-p}$ defines a 
cohomology class in $H^{6-p}(X;\IZ)$ which in turn is related to $H_{3+p}(X;\IZ)$ by Poincare duality. Hence,
$\sigma_{6-p}$  is related to the cycle $\Sigma_{3+p}$, where D-branes can also be wrapped.

Since this is the highest order differential map we will cosnider (in the presence of orientifolds, $d_5$ is in all cases trivial, although
in the abscence of them there are examples where $d_5$ is not; see \cite{JE1}), the algorithm ends at must at this step. 
So, K-theory classes are approximately\footnote{A K-theory group is equal to the Graded Complex $GrK(X)$ if the exact short 
sequence (see appendix B for notation)
\beqa
0\rightarrow K_p \rightarrow K_p/K_{p+1} \rightarrow K_{p+1} \rightarrow 0
\eeqa
is trivial for all $p$. This is not the case for $O3^-$-planes since the K-theory group related to this system is $KR^{-7}(\IRP^5)=\IZ$ while 
its Graded Complex is (as we shall see) $GrK(\IRP^5)=\IZ\oplus \IZdos$.}
given by the cohomology of $d_3$, and actually these classes must be classifying D-brane charges as well. So, it is important to give a
physical interpretation of closed and exact forms under $d_3$ in the context of D-branes.

\begin{figure}
\begin{center}
\centering
\epsfysize=7cm
\leavevmode
\epsfbox{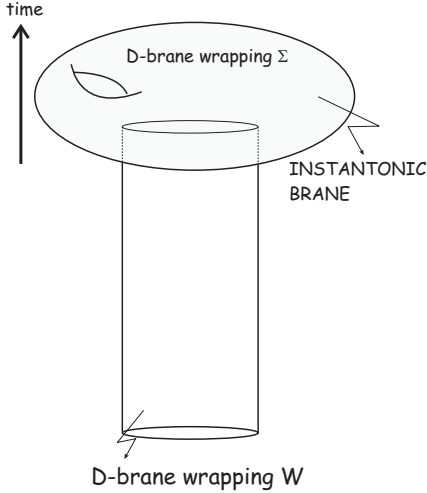}
\end{center}  
\caption[]{\small A $Dp$-brane wrapping the cycle ${\cal W}_p$ propagates in time and ends on an `instantonic' or `mortal brane' $D(p+2)$-brane wrapping a cycle $\Sigma_{p+3}$.}
\label{instantonic}   
\end{figure}

Closed forms, means that (see appendix B for terminology)
\beqa
d_3(\omega_{9-p}) = 0 = Sq^3(\omega_{9-p}) + [H_{NS}] \cup \omega_{9-p}\;,
\label{d3}
\eeqa
which actually is the condition to cancel anomalies as Freed and Witten showed in \cite{FW} (see also \cite{WB, Wk, DMW}. Then, closed forms under the differential map
$d_3$ are interpreted (via Poincare duality) as the allowed $p$-cycles where D-branes can be wrapped. In this context they are 
anomaly-free systems.
Now, let us turn our attention to exact forms. They can we expressed as
\beqa
d_3(\sigma_{6-p})= Sq^3(\sigma_{6-p}) + [H_{NS}]\cup \sigma_{6-p}=\omega_{9-p}=PD({\cal W}_p \subset
\Sigma_{p+3})\cup(\sigma_{6-p})\:.
\label{instanton}
\eeqa
This condition has a beautiful physical interpretation. D-branes can be wrapped on ${\cal W}_p$ since  such cycles satisfy the Freed-Witten
anomaly-free condition. However, they can be contained in higher cycles $\Sigma_{p+3}$ that do not satisfy the above anomaly-free condition
(since $d_3(\sigma_{6-p}) \neq 0$). Hence, D-branes wrapped on ${\cal W}_p$ can be, nevertheless unstable, if (\ref{instanton}) is satisfied.
The physical interpretation is that a brane wrapping a spatial cycle ${\cal W}_p$, propagates in time and terminates on a 
D-``instanton"\footnote{The term ``instantonic" refers to the fact that they are solutions of the equation of motion in either 
a Minkowski or Euclidean space and represent transitions between different sectors labeled by $D$-brane charges. In the Minkowskian case 
these branes involves tunneling. However, as was pointed out in \cite{JE3}, the term ``instantonic'' is not after all, related to a quantum 
theory. Rather, they are branes which are localized in time. For this reason, the author in ref. \cite{JE3} named these branes ``mortal branes".}
wrapping $\Sigma_{p+3}$. See figure \ref{instantonic}.

\subsection{Orientifold three-planes and the MMS picture}
\subsubsection{AHSS}
First of all, let us review the result given in \cite{BGS} from an alternative point of view (which is better for our purposes). Recall that
our goal is to interpret the equivalence of $\Oplus$ and $\Otplus$ in terms of instantonic branes.

As we saw in the previous section, by wrapping a $D3$-brane on a non-trivial zero-cycle of $\IRP^5$ we find the cohomology group which
classifies $D3$-branes on top of $O3$-planes (although this way to `justify' the presence of $D3$-branes on top of orientifolds seems to be
quite intricate, it will be very useful in the next examples). The cohomology group classifying such branes is $H^5(\IRP^5;\IZ)=\IZ$. This
is the group from which we start to construct our approximation to a K-theory group (actually the graded complex $GrK(X)$). So, let us apply
the differential\footnote{$\widetilde{d_3}$ is the twisted version of (\ref{d3}) and it takes twisted (untwisted) forms into untwisted
(twisted) ones. See appendix B.}
$\widetilde{d_3}$ to all non-trivial forms belonging to $H^5(\IRP^5;\IZ)$. Since $\widetilde{d_3}: H^5(\IRP^5;\IZ) \rightarrow H^8(\IRP^5,\ItZ)=0$, this map is
trivial. This means that the twisted Freed-Witten condition
\beqa
\widetilde{d_3}=\widetilde{Sq^3} +H_{NS} =0
\eeqa
is satisfied, and the system is non-anomalous (i.e., we can `wrap' $D3$-branes on zero cycles). From the mathematical point of view, this tells
us that five forms $[\omega_5]\in H^5$ are closed forms under $d_3$ ($d_3(\omega^5)=0$). Let us now study if there are five-forms which are
exact (in the context of MMS, this would mean that there are instantonic branes where a $D3$-brane wrapping a zero-cycle, could be unstable
to decay to vacuum). We would require then, a non-trivial map such that
\beqa
\widetilde{d_3}:H^2(\IRP^5,\ItZ) \rightarrow H^5(\IRP^5;\IZ)\;.
\eeqa
However, $H^2$ is trivial, and also this differential map. The conclusion is that there are not instantonic branes in this case.

Let us turn our attention to a $D5$-brane wrapping a non-trivial two-cycle. As we saw, this is actually a cohomology classification of
$O3$-planes. The first non-trivial approximation would be the existence of a non-trivial differential map, such that
\beqa
\widetilde{d_3}: H^3(\IRP^5;\ItZ) \rightarrow H^6(\IRP^5;\IZ)\;.
\eeqa
Again, since $H^6$ is zero, the map is trivial and the Freed-Witten condition to cancel anomalies is satisfied. This is:
\beqa
\widetilde{d_3}(G_3)\;=\;\widetilde{Sq^3}(G_3) +H_{NS}\cup G_3 =0
\eeqa
Notice that in a homology description the above condition is given by $W_3(\IRP^2) +H_{NS}|_{\IRP^2}=0$, where $\IRP^2$ is the two-cycle where a
D5-brane is wrapped. The physical interpretation is that a $D5$-brane is allowed to wrap $\IRP^2$ in the presence of an $O3$-plane, or equivalently
that $G_3$ is a closed three-form under $\widetilde{d_3}$.
However, this brane could be unstable to decay into vacuum if there exists an instantonic brane where the brane ends. The existence of such
instantonic brane would be given if $G_3$ were an exact form. Hence we must look for non-trivial and surjective $\widetilde{d_3}$ maps, such that
\beqa
\widetilde{d_3} : H^0(\IRP^5;\IZ) \rightarrow H^3(\IRP^5,\ItZ)\;.
\eeqa
In fact, since $H^0=\IZ$, the above map is non-trivial under certain conditions which involve the presence of a non-trivial $H_{NS}$-flux.
As we shall soon see, $\widetilde{Sq^3}(\eta_0)=0$, and the map is given by
\beqa
\widetilde{d_3}(\eta_0)\;=\;H_3 \cup \eta_0\;.
\eeqa
It is easy to read off from this expression that only in the presence of a non-trivial flux $H_{NS}$, the differential map $\widetilde{d_3}$ will be
non-trivial. However, $H_{NS}\in \widetilde{H^3}=\IZdos$ is actually the cohomology class which classifies $\Ominus$ and $\Oplus$ -planes. Then, the above
map is trivial if the orientifold we are considering is $O3^-$ and non-trivial if it is $O3^+$ The latter case means that the field strength
$G_3$ is an exact form given by $G_3=\widetilde{d_3}(\eta_0)$. 

Before given a physical interpretation of this result, let us point out some observations:
1) From the above results, we conclude that
\beqa
\begin{array}{rcl}
Ker\; \widetilde{d_3}&=&\IZ\oplus\IZdos\\
Im\; \widetilde{d_3}&=& 
\left\{
\begin{array}{c}
0 \quad \hbox{for $\Ominus$}\\
\IZdos \quad \hbox{for $\Oplus$}
\end{array}
\right.
\end{array}
\eeqa
Then, since the graded complex is given as $GrK(\IRP^5)= Ker\;\widetilde{d_3}/Im\;\widetilde{d_3}$, we conclude that $GrK(\IRP^5)=\IZ$ for an $\Oplus$ and to $\IZ\oplus \IZdos$ for an
$\Ominus$. Notice also that the latter one differs from the K-theory value given in \cite{BGS} (see footnote in section 3.1). This is for the non-triviality of the
related spectral sequence which actually explains the relative charge between $\Ominus$ and $\Otminus$.

2) It is followed that in the presence of a non-trivial $H_{NS}$ ($\Oplus$ or $\Otplus$), $G_3$ is lifted to the 
zero-class in the Graded Complex, and also in the K-theory group (since in this case both are equal). Since $[G_3]\in\IZdos$ classifies $\Oplus$ and $\Otplus$, these two
orientifolds are equivalent in the K-theory classification (both of them are lifted to the same trivial class). Notice also that until now,
we are only classifying RR fields. The information concerning the NS-NS field $H_{NS}$ is inserted via the differential map $\widetilde{d_3} = \cup H_3$.\\

3) As was pointed in \cite{BGS}, we can identify $\eta_0$ with the RR field $C_0$. This is achieved from the above non-trivial map. Under $\widetilde{d_3}$,
even integers in $H^0$ are sent to the zero class in $\widetilde{H^3}=\IZdos$ while odd integers are sent to the non-trivial class $[1]\in \IZdos$. In other
words, even-integers corresponds, via $\widetilde{d_3}$, to an $\Oplus$-plane, while odd-integers are related to $\Otplus$. If we shift $\eta_0$ by one
unit, $\eta_0 \rightarrow \eta_0 +1$, we will get the transformation $\Oplus \rightarrow \Otplus$.  So, it is natural to identify $\eta_0$ with
$C_0$ which in turn is identified with the $\theta$-term in the complex coupling in four- dimensional ${\cal N}=4$ SYM theories. 

\subsubsection{The MMS-picture}
\begin{figure}
\begin{center}
\centering
\epsfysize=7cm
\leavevmode
\epsfbox{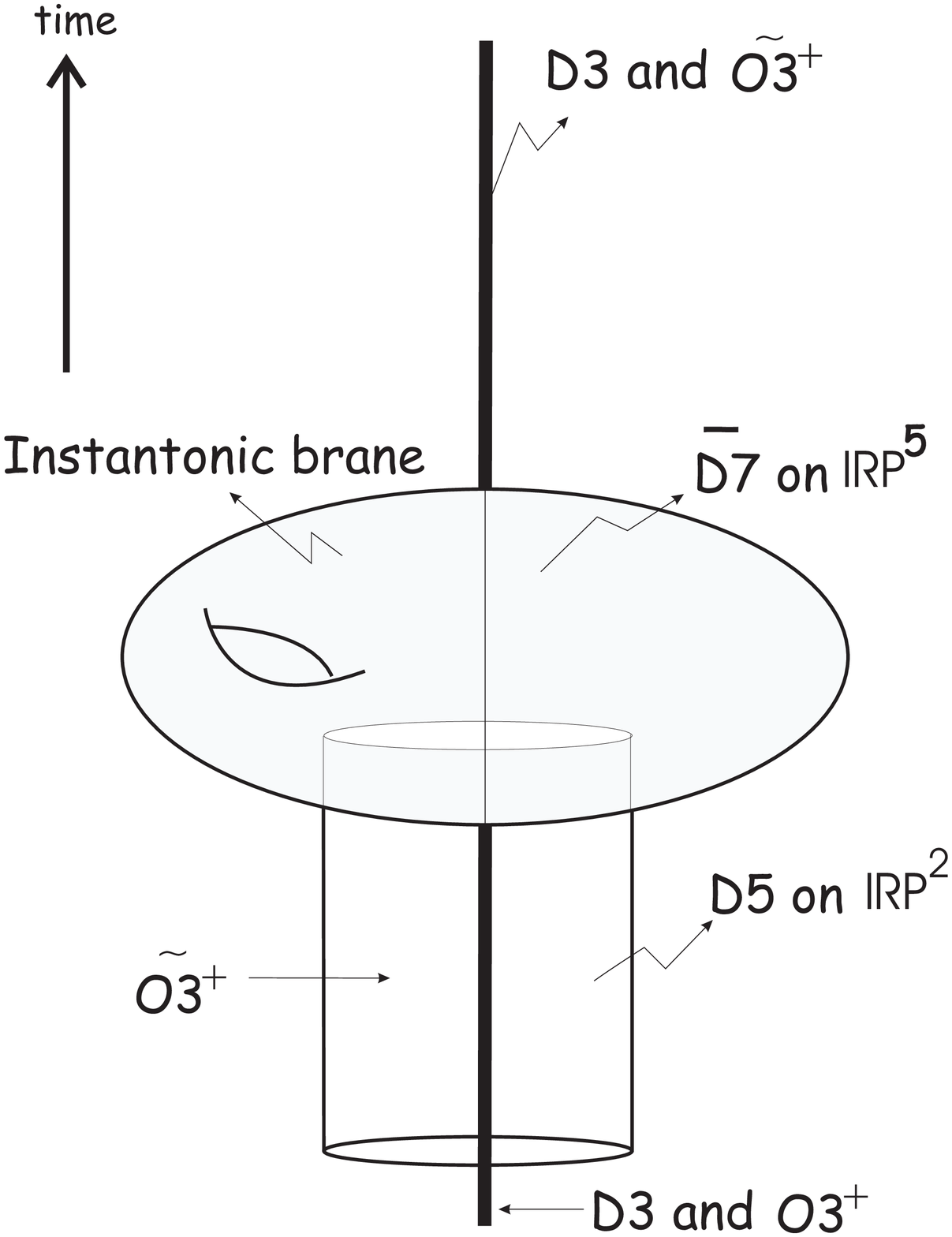}
\end{center}  
\caption[]{\small A $D5$-brane wrapping $\IRP^2$ transverse to an $O3^+$-plane is unstable to decay into vacuum in the presence of an 
instantonic or mortal $\Dseven$-brane wrapping $\IRP^5$. Since $\Otplus$ is constructed by an $\Oplus$ and a $D5$-brane on $\IRP^2$, 
we conclude that an $\Otplus$ transforms into an $\Oplus$ via the interaction with the seven-brane.}
\label{o3equivalence}   
\end{figure}

It is time to explain the physical interpretation of $\eta_0$ in the context of the MMS-picture (although we already have seen
that this form is actually the RR field $C_0$). As we saw at the beginning of this section, we can relate differential forms with homology
cycles. In our case, $H^0\sim H_5$ which gives us five-cycles where we can wrap $D$-branes. Also, we saw that the only brane we can
wrap on an untwisted five-cycle is a $D7$-brane. Hence, this means that a $D7$-brane wrapping $\IRP^5$ represents an instantonic or mortal 
brane where
a $D5$-brane wrapping $\IRP^2$ (in the presence of an $O3^+$-plane) is unstable to decay into vacuum\footnote{In order to prevent divergences
in energy in the transversal non-compact coordinate to the seven-brane, we restrict our picture as a local model}. An schematic picture is 
given in figure \ref{o3equivalence}.

Some comments are in order: a) Actually,
for consistency the seven-brane is an anti-$D7$-brane (labeled as $\Dseven$). See Appendix B. b) This picture is in accordance with the
identification of $\eta_0$ as the RR field $C_0$ since a $\Dseven$-brane is magnetically charged under $C_0$. And finally c) $\IRP^5$ is an
orientable manifold and it is spin$^c$ (see \cite{FW}). This means that $\widetilde{Sq^3}(\eta_0)=W_3(\IRP^5)=0$. Then, the action of the differential map
$\widetilde{d_3}$ over $C_0$ is 
\beqa
\widetilde{d_3}C_0 =C_0\cup H_{NS}=G_3
\label{mera}
\eeqa

The above picture can be interpreted in different but nevertheless, related ways. 
A $D5$-brane wrapping $\IRP^2$ is unstable to decay into vacuum by the presence of an instantonic $\Dseven$-brane wrapping $\IRP^5$ transverse to
an $\Oplus$-plane. However since a $D5$-brane wrapping $\IRP^2$
transverse to an $\Oplus$ -plane is actually the cohomology version of an $\Otplus$-plane, this is unstable to decay
into an $\Oplus$ in the presence of the instantonic seven-brane.

The above configuration of branes in the covering space is as follows\footnote{Notice that this model does not T-dualize to a stable object
in type $USp(32)$ theory. In fact, the seven-brane T-dualizes to a $D3$-brane in ten-dimensions, which is not classified by K-theory. The
only three-dimensional object classified by K-theory in type $USp(32)$ theory is a
non-BPS state, which is constructed by the pair $D3-\ov{D3}$. Then, the $\Dseven$ -brane considered in our model must not be classified by
K-theory since it is only a `cohomological' object. Indeed, according to equation (\ref{mera}) this seven-brane is obstructed to be lifted to
K-theory since is ts represented by an exact form under $\widetilde{d_3}$. Notice also that the image of $\Dseven$ under the presence
of an orientifold three-plane, is also a $\Dseven$. This is due to the fact that type IIB RR field $C_8$ is invariant under the projection of
the orientifold three-plane. After wrapping the seven-brane on the non-trivial five-cycle, we get an integer-charged object. This also
contradicts the fact that stable objects in type $USp(32)$ related by T-duality to these seven-branes, must have discrete topological charge
$\IZdos$. The reason for that is that wrapping a seven-brane on $\IRP^5$ give us a cohomological version of this objetc which must be refined
by a K-theory one. Finally, we  do not care about cancelation of tadpoles (global gauge anomalies in lower dimensional probe branes)
comming from the compactification of type $USp(32)$ theory, as in \cite{urangak}, since one transversal coordinate to the seven brane is
non-compact.}:
\begin{center}
\begin{tabular}{ccccccccccc}
&0&1&2&3&4&5&6&7&8&9\\
$\Dseven$&$-$&$-$&$\cdot$&$-$&$-$&$-$&$-$&$-$&$-$&$\cdot$\cr
$\Oplus$&$-$&$-$&$-$&$-$&$\cdot$&$\cdot$&$\cdot$&$\cdot$&$\cdot$&$\cdot$\cr
$D3$&$-$&$-$&$-$&$-$&$\cdot$&$\cdot$&$\cdot$&$\cdot$&$\cdot$&$\cdot$\cr
$D5$&$-$&$-$&$-$&$-$&$-$&$-$&$\cdot$&$\cdot$&$\cdot$&$\cdot$
\end{tabular}
\end{center}
Once we put time on coordinate two, the $\Dseven$-brane wraps on coordinates 45678 on the
covering space (on $\IRP^5$ after taking the orientifold projection) and it is `instantonic' to the transverse space of $\Oplus$. Notice also
that this implies that the five-brane is wrapping coordinates 45 in the covering space and $\IRP^2$ after the orientifold projection.          

Before concluding this section, let us give a couple of arguments which support the consistency of the instantonic seven-brane picture.\\

\noindent
{\bf SUGRA Bianchi identities}

Here we want to point out that the above results can also be inferred from the Bianchi identities in supergravity IIB. We closely follow the
ideas given in \cite{DMW, JE3, JE1, JE2}.

As we saw, a $\Dseven$-brane wrapping $\IRP^5$ is an anomalous system since the condition to avoid anomalies is not satisfied. This
inconsistency (an ill-defined sign for the pfaffian in the path integral) is cured by adding magnetic sources to the worldvolume of the
seven-brane, which are $D5$-branes wrapping $\IRP^2$.

The same results are gathered by studying the SUGRA ten-dimensional Bianchi identities. Consider a $\Dseven$-brane wrapping $\IRP^5$ such
that there is a subcycle $\IRP^3\subset\IRP^5$ supporting $k$-units of $H_{NS}$-flux. The presence of a $\Dseven$-brane induces a non-trivial
$G_1$-flux. Then the product $G_1\cup H_{NS}$ must be different from zero. According to the Bianchi identities \cite{sugraeqns}, 
\beqa
dG_3=G_1\cup H_{NS}
\label{bianchi}
\eeqa
and then, $dG_3\neq 0$. This equation stands for a current of $D5$-branes emanating from the seven-brane, perpendicular to the three-cycle $\IRP^3$ which support
the $k$-units of $H$-flux (coordinates 456 in the covering space) and to the seven brane (coordinate 9 in covering space). Hence, by
construction, we have a current of $D5$-branes wrapping a two-cycle (coordinates 78). Hence, we require the presence of these five-branes in
order to have a consistent system. Notice that $C_0\cup H_{NS}\;=\;G_3$  implies equation (\ref{bianchi}) by
applying $\widetilde{d_3}$ to $C_0$, and also that $G_3$ is an exact form under $d$. This in turn means that $G_3$ is lifted to a trivial class in
K-theory. 
Notice
also that the equation (\ref{bianchi}), implies the violation of the current conservation since
\beqa
d\ast j_6\,=\,\ast j_8\cup H_{NS}
\eeqa
which in turn tell us that $D5$-branes must be going out from the $\Dseven$-brane. Two comments are in order:
\begin{enumerate}
\item
By integrating over two sides of (\ref{bianchi}) we get,
\beqa
Q_{D5}\,=\,\int_{\IRP^4} dG_3\,=\,\int_{\IS^1} G_1 \cdot \int_{\IRP^3} H_{NS}\,=\,k\;.
\eeqa
Notice also that in our case, since $H_{NS}$ is related to the presence of an $\Oplus$-plane, $k$ is actually an integer mod 2 (or a
half-integer mod 1), and then the current of $D5$-branes must have a $\IZdos$-charge. But this is actually what we had previously: a
$D5$-brane wrapping $\IRP^2$ acquires a discrete torsion charge since a two-cycle is classified by $H_2(\IRP^5;\ItZ)=\IZdos$.
Again, by interpreting a $D5$-brane wrapping  $\IRP^2$ in the presence of an $\Oplus$-plane as an $\Otplus$-plane, what we have is that a
$\Dseven$-brane wrapping $\IRP^5$, transverse to an $\Oplus$-plane, induces a `current of an $\Otplus$-plane'.
\item
$H_{NS}$ can be interpreted as a current of a four-dimensional object on the worldvolume of a $\Dseven$-brane. In fact, since a five-brane is
intersecting $\Dseven$ in a four-dimensional spatial submanifold (as in our model) there is a current in the worldvolume $W_8$ of the $\Dseven$ given
by $\ov{\ast}\ov{j_5}$, where $\ov\ast$ is the Hodge dual operator which takes $p$-forms into $(8-p)$-forms in $W_8$, and $\ov{j_5}$ is the
current associated to the four-dimensional object. It can be shown (see \cite{Be}) that
\beqa
d\ast j_6 \,=\,\ast j_8 \cup \ov{\ast} \ov{j_5}
\eeqa
where $\ast j_6$ is the current associated to the emerging $D5$-brane and $\ast j_8$ to th $\Dseven$-brane.

Since we are working in type IIB theory, topologically a $D5$ and a $NS5$ -branes can be expressed in similar six-forms. According to our
model, $\ov{\ast}\ov{j_5}$ can not be associated to a $D5$-brane since the presence of the latter is precisely a consequence of having a
non-trivial $\ast j_6$ current. Hence it must be related to a $NS5$-brane. But a $NS5$-brane (that in our configuration is wrapping
$\IRP^2$) is the one which give us a non-trivial value of $H_{NS}$. Then the picture matches completely: a current of $D5$-branes is followed by
a non-trivial value of $H_{NS}$ which in turn is a consequence of wrapping $NS5$-branes on $\IRP^2$ and intersecting the $\Dseven$ in
a four-dimensional submanifold.
\end{enumerate}

\noindent
{\bf A tadpole in the field content}.

Here we follow closely the argument given in the Appendix A in \cite{urangac}. The question is: what is the consequence of such inconsistency in the
equations of motion for  fields living on the worldvolume of the seven-brane? (in the absence of the five-brane). The answer is that a
tadpole appears and it can be removed once we add five-branes wrapping a two-cycle of $\IRP^5$.

Let us say we have a $\Dseven$-brane wrapping the five-cycle $\IRP^5$. In the presence of a non-trivial $H$-field
there is a coupling (in the worldvolume of a $D7$-brane), given by
\beqa
\int_{\IRP^5\times\IR^{1,2}}H_{NS}\times\widetilde{A}_5\;=\;(\frac{1}{2}\;mod\;1)\int_{\IRP^2}\widetilde{A}_5
\eeqa
where clearly we have a tadpole. However, we can remove this tadpole by adding a magnetic source for $\widetilde{A}_5$. This is given by a
$D5$-brane in coordinates 012378. The $D5$-brane is a four-dimensional object in the perspective of the world-volume of the seven-brane
(which is in coordinates 01345678). Hence, by adding $(\frac{1}{2} \;mod\;1)$ units of a $D5$-brane wrapping coordinates 89, we remove the
tadpole since the action becomes
\beqa
S\sim \int G_6\wedge \widetilde{G_6}\;+\; (\frac{1}{2}\;mod\;1)\int\widetilde{A_5}\;,
\eeqa
which in turns implies that the charge of the four dimensional object in the perspective of the seven-brane worldvolume, must be a
half-integer mod 1, which is actually the one obtained by wrapping a $D5$-brane on $\IRP^2$.

\subsection{$\cal T$-duality for ${\cal N}=4$ SYM}
Here, we study the effect of the instantonic seven-brane in the field content on $D3$-branes on top of $\Oplus$-planes. The constant coupling
in Type IIB string theory,
\beqa
\tau\;=\;C_0\;+\;\frac{i}{g_s}\;,
\eeqa
is non-trivial under the presence of a seven-brane, since the charge of the seven-brane is measure by its coupling with the complex dilaton,
\beqa
Q_{\Dseven}\;=\ \partial(C_0 +ig^{-1}_s)\;=\;\Delta C_0\;,
\eeqa
which in turn implies that
\beqa
C_0 \rightarrow C_0 + Q_{\Dseven}\;.
\eeqa
Hence, in the presence of a single anti-seven-brane there is a shifting of the RR field $C_0 \rightarrow C_0 -1$. Notice that although the
seven-brane is instantonic in the sense that is localized in time, after it disappears (it is a `mortal brane') it leaves a remanent 
unit of $C_0$-flux which give us the above shifting. Of course, as it is
well-known, this implies a shifting in the complex coupling $\tau$:
\beqa
\tau \rightarrow \tau -1\;.
\eeqa
Such a shifting is also reflected in the four-dimensional ${\cal N}=4$ SYM theory, where $\tau = \frac{\theta}{2\pi}+ \frac{4\pi i}{g^2_{SYM}}$. 

 Also we can find the transformation in $\tau$ only by looking at equation (\ref{mera}). This equation tells us that for each
 non-trivial unit of $H_{NS}$-flux, we get a non-trivial unit of $G_3$. If we array the fluxes $G_3$ and $H_{NS}$ as a doublet, this means that
\beqa
\left(
\begin{array}{c}
[G_3]\\
\lbrack H_{NS}\rbrack
\end{array}
\right) \rightarrow \left(
\begin{array}{c}
[G_3]+[H_{NS}]\\
\lbrack H_{NS}\rbrack
\end{array}
\right)\;.
\eeqa
 Now, since $d_3$ is a linear map, we expect that the pair $([G_3],[H_{NS}])$ transforms as a doublet under the action of a $2\times 2$
 matrix,
 \beqa
 \left(
 \begin{array}{c}
 [G_3]+[H_{NS}]\\
 \lbrack H_{NS}\rbrack
 \end{array}
 \right)\,=\,\left(
 \begin{array}{cc}
 p&r\\
 q&s
 \end{array}
 \right)
 \left(
 \begin{array}{c}
 [G_3]\\
 \lbrack H_{NS}\rbrack
 \end{array}
 \right)\;,
 \eeqa
 where in principle there are not restrictions to the values of the matrix entrees. Then we get that
 \beqa
 \left(
 \begin{array}{cc}
 p&r\\
 q&s
 \end{array}
 \right)\;=\;\left(
 \begin{array}{cc}
 1&1\\
 0&1
 \end{array}
 \right)
 \eeqa
 Now, since $\widetilde{d_3}:\IZ \rightarrow \IZdos$, a shift in $C_0 \rightarrow C_0 -1$ implies a shift in $G_3$ from the trivial to the non-trivial
 class (or viceversa). Physically this is the interchange between $\Oplus$ and $\Otplus$ (as was found in \cite{BGS}). Hence, by only studying
 the relation (\ref{mera}) we conclude that the above two orientifolds are related by a transformation in $\tau \rightarrow \tau -1$, which as
 we already know, is the ${\cal T}\in SL(2,\IZ)$-duality. The shif in $C_0$ (theta angle) gives all dyons an extra unit of electric charge,
 swaping the symmetry group of monopoles and dyons in accordance with table 2.

However we
want to stress that in this picture, such a duality is a consequence of the topology of the space in ten-dimensions. This is, the presence of
the instantonic seven-brane is completely determined by the AHSS. In the next section, we shall see that we can obtain the full symmetry by
considering other instantonic branes.

Notice also, that the four-dimensional ${\cal N}=4$ SYM theory with gauge group $SO(2N)$ is not affected by the presence of an instantonic
brane since in the presence of an  $\Ominus$-plane, no $D5$-branes are unstable to decay into vacuum (or to be created) after the interaction
with the former one. This implies that such a theory is self-dual, at least, under ${\cal T} \in SL(2;\IZ)$ duality.

\section{$SL(2,\IZ)$-duality from instantonic seven-branes}
Now, we want to extend this result in order to get that the full duality under $SL(2, \IZ)$ is found via the presence of instantonic
branes. In particular, we want to give a picture of the strong-weak duality, which transforms an $\Oplus$ into an $\Otminus$ (and viceversa).
However we require a proper classification of RR fields as well as NS-NS fields. We already have done the latter since we know how to
classify the field strength $H_{NS}$ in cohomology, but we require a version of AHSS which could also take into account the presence of
NS-NS-fields.

In this section we give a more general picture of
instantonic branes which lead us to the conclusion that, under this general classification of RR and NS-NS-fluxes by cohomology and their
lifting to K-theory, a $SL(2,\IZ)$-duality must be present in ${\cal N}=4$ SYM theories with different gauge groups.

\subsection{NS-NS fluxes and the AHSS}
Up to now, we know that we can wrap D-branes on non-trivial cycles ${\cal W}$ if the Freed-Witten condition is satisfied:
\beqa
W_3({\cal W}) + [H_{NS}]|_{\cal W}=0\;.
\eeqa
Its cohomology version reads,
\beqa
\widetilde{d_3}(\omega_p)= \widetilde{Sq^3}(\omega_p) +H_{NS}\cup \omega_p =0\;.
\eeqa
What we are doing here is to compute the image of a RR-field $\omega_p$ (which is related to an $(n-p)$-cycle ${\cal W}$ where a $D$-brane is
wrapped) under the differential map $\widetilde{d_3}$. The next step is to look for an expression, similar to $\widetilde{d_3}$, 
which classifies NS-NS fluxes as well.

In our case this can be achieved by taking advantage of the fact that $G_3$ and $H_{NS}$ are both classified by $H(\IRP^5;\ItZ)$. From the
topological point of view we can invert the roles played by $G_3$ and $H_{NS}$ without altering the mathematical map $\widetilde{d_3}$. 
So we have at
least two interesting cases, which read
\beqa
\widetilde{d_3}(C_0)=C_0 \cup H_{NS} = G_3\\\nonumber
\widetilde{d_3}(\widetilde{C_0})=\widetilde{C_0} \cup G_3 =H_{NS}\;,
\eeqa
where $\widetilde{C_0}$ is a scalar field (zero-form) different to $C_0$.
The first of the two above equations has already a physical interpretation as was seen in the previous section. However the last one 
is physically different since
now we are classifying NS-NS fluxes instead of RR ones. This is the version of the linear map $\widetilde{d_3}$ we were looking for. 
Before gathering all
the physics that the last linear $\widetilde{d_3}$ map implies, it is important to point out that a similar expression was proposed in
\cite{JE1}, 
where the
authors used S-duality in type IIB string theory in order to construct an AHSS version which also classifies NS-NS fluxes. Although this work
was motivated by their result, in our case we are working only with 3-dimensional objects embedded in a ten-dimensional spacetime that allow
us to not consider S-duality as the way to built up a general AHSS, while in
\cite{JE1} the authors proposed an AHSS algorithm which can be applied to different objects.

Let us study the map $\widetilde{d_3}$ which classifies NS-NS fluxes
(from now on, we will label $\widetilde{d^{NS}_3}$ the map which takes NS-NS forms, and $\widetilde{d^{R}_3}$ the one
which takes RR forms),
\beqa
\widetilde{d^{NS}_3}(\omega_p^{NS})= \widetilde{Sq^3}(\omega^{NS}_p) + G_3\cup \omega^{NS}_p =0\;.
\eeqa
This means that the condition to wrap NS-NS charged-objects (as a $F1$-string, $NS5$-brane or the S-dual of seven-branes) is that, 
in the absence of a RR flux $G_3$
a NS-NS object must be wrapped on Spin$^c$-manifolds (see \cite{JE1}). If the manifold is
 Spin$^c$ but there is a non-trivial $G_3$
flux, the condition now reads that $G_3\cup\omega_p$ must belong to the zero class in cohomology. If such a condition is not satisfied, then
the system is said to be anomalous.

Applying this proposal to our picture of orientifold three-planes, there is a very important change, since now, the Freed-Witten condition
for NS-NS-fluxes to cancel anomalies depends on the presence of a non-trivial $G_3$-flux, which is related to $\Otplus$ and $\Otminus$-planes.
Hence, following the same arguments than in the previous section, the only non-trivial $\widetilde{d^{NS}_3}$ map is
\beqa
\label{sfw}
\widetilde{d^{NS}_3}: H^0(\IRP^5;\IZ) \rightarrow H^3(\IRP^5;\ItZ)\\\nonumber
\widetilde{d^{NS}_3}(\widetilde{C_0)}=G_3\cup \widetilde{C_0}= H_{NS}
\eeqa
Notice that now, $H_{NS}$ is a closed form under $\widetilde{d^{NS}_3}$ if $G_3$ is non-trivial and then, it must be lifted to a trivial class in the correspondent K-theory
classification (a twisted K-theory). This means that the two orientifolds classified by $\widetilde{H^3}$ (and remember
that now we are classifying NS-NS fluxes) $\Otplus$ and $\Otminus$ ($G_3$ is non-trivial!) are equivalent in this topological
classification.

\subsection{Physical meaning of $\widetilde{C_0}$}
Although we do not require to know exactly what it is the seven brane related to $\widetilde{C_0}$ in equation (\ref{sfw}), we indeed can elucidate
its nature. For that we use S-duality\footnote{Notice that we are using S-duality only to know what the seven-brane related to
$\widetilde{C_0}$ is. Remember that our goal is to find S-duality as a consequence of the map (\ref{sfw}).}.
Since equation (\ref{sfw}) is obtained by interchanging $H_{NS} \leftrightarrow G_3$, we expect that $\widetilde{C_0}$ must be coupled to 
the S-dual version of a $D7$-brane.

S-dual versions of $D7$-branes, or more generally, $SL(2,\IZ)$-versions, were studied for instance in \cite{sevenb1, sevenb2, sevenb3}. In general, a $(p,q)$-seven
brane is obtained after applying a transformation 
\[
\left(
\begin{array}{cc}
p&r\\
q&s
\end{array}
\right)
\]
in $SL(2,\IZ)$ to a $D7$-brane, where the pair $(p,q)$ transforms as a
doublet under this map. A $(p,q)$ seven-brane is the endpoint of $(p,-q)$ dyonic five-branes and $(p,q)$- dyonic strings and in this context,
a $D7$-brane is a $(1,0)$ seven-brane and its S-dual version ($p=s=0$ and $q=-r=1$) a $(0,1)$ seven-brane. The scalar field $C_0$ 
(a RR field)
couples magnetically to a $D7$-brane while a $(0,1)$ seven-brane couples with the S-dual version of $C_0$ which is actually a scalar field
where the dilaton has been turned on. We denote this field as $\widetilde{C_0}$, where
\beqa
\widetilde{C_0}= \frac{-C_0}{C_0^2 + e^{-2\phi}}\;.
\eeqa

Also notice that in this case, $\widetilde{H^3} \sim \widetilde{H_2}$ is classifying $NS5$-branes which wrap non-trivial two-cycles of $\IRP^5$.
Hence, the picture we get from equation (\ref{sfw}) is that in the presence of an instantonic $(0,-1)$ seven-brane which wraps $\IRP^5$
transverse to an $\Otminus$-plane, a $NS5$-brane wrapping $\IRP^2$ is unstable to decay into vacuum via the interaction with the seven-brane.
This also means that in the presence of such instantonic $(0,-1)$-brane, an $\Otplus$ transforms into a $\Otminus$-plane.

The resulting picture involves a change in the complex coupling, since it is not trivial under the presence of a generic $(p,q)$ seven-brane.

So, according to this picture, by classifying NS-NS fluxes, an $\Otminus$-plane becomes an $\Otplus$-plane via the interaction with an
instantonic $(0,-1)$ seven-brane wrapping $\IRP^2$ transverse to the orientifold. The seven-brane in turns, must establishes a non-trivial
monodromy in $\tau$. Although this monodromy is well known and also can be derived by using S-duality, here we shall show that without using
S-duality, we can obtain the rule transformation in $\tau$ as a result of the presence of the instantonic $(0,-1)$-seven-brane.

Equation (\ref{sfw}) tell us that for each non-zero unit of $G_3$ flux, we get a unit of $H_3$-flux. Hence, according to this rule and
following the same reasoning as in section 2, we have that
\beqa
\left(
\begin{array}{c}
[G_3]\\
\lbrack H_{NS}\rbrack
\end{array}
\right) \rightarrow 
\left(
\begin{array}{c}
[G_3]\\
\lbrack H_3\rbrack +\lbrack G_3\rbrack
\end{array}
\right)\;=\;
\left(
\begin{array}{c}
[G_3']\\
\lbrack H_3'\rbrack
\end{array}
\right)\;,
\eeqa
where
\beqa
\left(
\begin{array}{c}
[G_3']\\
\lbrack H_3'\rbrack
\end{array}
\right)\;=\;
\left(
 \begin{array}{cc}
 p&r\\
 q&s
 \end{array}
 \right)
\left(
\begin{array}{c}
[G_3]\\
\lbrack H_{NS}\rbrack
\end{array}
\right) \;.
\eeqa
Then, we get that
\beqa
\left(
 \begin{array}{cc}
 p&r\\
 q&s
 \end{array}
 \right)\;=\;
 \left(
 \begin{array}{cc}
 1&0\\
 1&1
 \end{array}
 \right)\;:=\;{\cal U}\;,
\eeqa
and by similarity with the picture given in previous section, we also conclude that under the presence of the seven-brane, the complex
coupling is also transformed by the above matrix, i.e.,
\beqa
\tau \rightarrow \frac{p\tau + r}{q\tau +s}=\frac{\tau}{\tau +1}
\eeqa
A comment is given to support this idea: Since our picture involves a $NS5$-brane ending on a seven-brane, this brane must be actually a
$(0,-1)$ seven-brane, a result which matches with our previous statement using S-duality. The monodromy in $\tau$ under a $(0,-1)$ seven
brane is easily computed, giving us the same result as above. Even more, notice that the equivalences $\Oplus \sim \Otplus$ and $\Otplus \sim
\Otminus$ give us two pictures (in terms of instantonic branes) which are related each other by an S-dual transformation, confirming that our
conclusions are well-supported by very well-known facts.

\begin{figure}
\begin{center}
\centering
\epsfysize=7cm
\leavevmode
\epsfbox{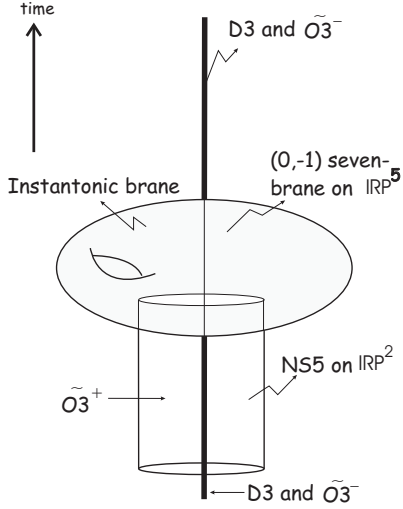}
\end{center}  
\caption[]{\small An $\Otplus$-plane transforms into an $\Otminus$ in the presence of an instantonic or mortal $(0,-1)$ seven-brane wrapping
$\IRP^5$.
Equivalently, a $NS5$-brane wrapped on $\IRP^2$ is unstable to decay into vacuum via the presence of the seven-brane and a non-trivial RR
$G_3$-flux.}
\label{NSo3dualities}   
\end{figure}

In this sense, we obtain that orientifolds $\Otplus$ and $\Otminus$ are related each other by an ${\cal
U}$-transformation, which indeed is correct as was shown in section 1. The picture is schematized in figure \ref{NSo3dualities}.\\

\noindent
{\bf SUGRA Bianchi Identities}

In contrast to the case of a $D5$-brane emanating from a $\Dseven$-brane wrapping $\IRP^5$, in this case the Bianchi identity to be
considered is 
\beqa
dH_{NS}=0
\eeqa
since $H_{NS}=dB_{NS}$. This identity does not imply a flux of $NS5$-branes from the seven-brane wrapping a cycle which
supports a non-trivial amount of $G_3$-flux. However we can modify the above identity to $dH_{NS}=\ast j^6_{NS}$ to allow the presence of
$NS5$-branes, as far as locally we keep $H=dB$. This is indeed the difference between the case in section 2 and the present one: here, the
Bianchi identity must be modify in order to gather a flux of $NS5$-branes.

We propose that a source of $H_{NS}$ is also given by the product of fields $\widetilde{G_1}\cup G_3$, where $\widetilde{G_1}=d\widetilde{C_0}$.
Then, we have
\beqa
dH_{NS}\;=\;\widetilde{G_1}\cup G_3
\label{sugrans}
\eeqa
which tell us that a $NS5$-brane must be emanating from the $(0,-1)$ seven-brane. By
integrating over two sides of the above equation, we get that the charge of the $NS5$-brane must be equal to $k$. 
Notice that, in a similar fashion with section 1, we get,
\begin{enumerate}
\item
The charge carried by the $NS5$-brane must be $\IZdos$ since $[G_3]\subset\IZdos$ and then $\int_{\IRP^3}G_3=\frac{1}{2}\;mod\;1$. A
non-trivial value of $G_3$ stands for an $\widetilde{O3}$-plane. This matches with the fact that the $NS5$-brane is wrapping $\IRP^2$, which 
belongs
to $\IZdos$ and that actually is the cohomology version of the orientifold $\widetilde{O3}$.
\item
The source term introduced above, also implies a violation in the current conservation, since now
\beqa
d\ast j_6^{NS}\;=\;\ast\widetilde{j_8}\cup G_3
\label{vcns}
\eeqa
where $j^{NS}_6$ represents the current of $NS5$-branes and $\widetilde{j_8}$ is the current associated to the seven-branes. Such a
violation in the current conservation means that there is a current of $NS5$-branes emanating from the seven-branes. Notice also that we can
write equation (\ref{vcns}) as
\beqa
d\ast j_6^{NS}\;=\;\ast\widetilde{j_8}\cup \ov\ast{\ov{j_5}}\;,
\eeqa
where $\ov{j_5}$ is the current associated to a four dimensional object lying on the worldvolume of the seven-brane. In fact this current is
associated to $D5$-branes wrapping $\IRP^2$ around an orientifold three plane and intersecting the seven-brane in a four-dimensional
submanifold, which give us the non-trivial values of the flux $G_3$. So
indeed, $G_3$ can be also represented by the above current and expression (\ref{sugrans}) makes sense.
\end{enumerate}

\subsection{Generating the full $SL(2,\IZ)$ duality}
\begin{figure}
\begin{center}
\centering
\epsfysize=7cm
\leavevmode
\epsfbox{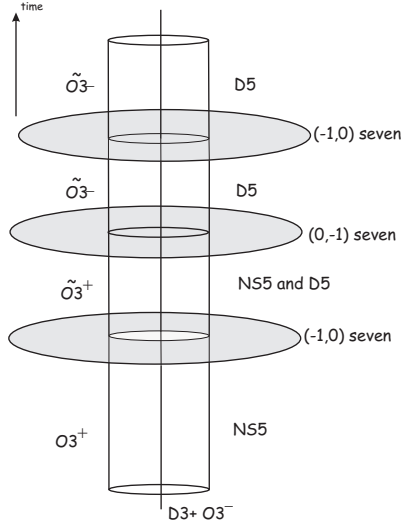}
\end{center}  
\caption[]{\small An $\Oplus$-plane is transformed into an $\Otminus$ in the presence of three instantonic or mortal seven-branes. These
branes in turn, generate the $\cal S$-duality in the field content on the worldvolume of $D3$-branes on top of the orientifold planes.}
\label{U1}   
\end{figure}

Altogether, the above two pictures generate the full $SL(2,\IZ)$ symmetry, since by the presence of an instantonic
$\Dseven$ -brane we have the transformation $\tau \rightarrow {\cal T}^{-1} \tau$ while in the presence of an instantonic $(0,-1)$
seven-brane, $\tau \rightarrow {\cal U}\tau$.  So, if we select these two transformations as the generators of the full symmetry, any type of
transformation $\tau \rightarrow \frac{p\tau +r}{q\tau +s}$ is generated by a chain of ${\cal T}^{-1}$'s and ${\cal U}$'s. In particular we are
able to induce the presence of the strong-weak duality (S-duality) by a chain of instantonic branes intersected by $D3$-branes and
orientifold three-planes. Since
\beqa
S\;=\;{\cal T}^{-1}{\cal U}{\cal T}^{-1}
\eeqa
we conclude that three instantonic seven-branes, given by a $\Dseven$, a $(0,-1)$ and a $\Dseven$ -branes (in that precisely order)  generate
S-duality in the field content on the worldvolume of $D3$-branes on top of orientifold three-planes. According to section 1, we already know
that under S-duality, an $\Oplus$ -plane is transformed into an $\Otminus$ one, while $\Otplus$ is invariant. Let us see if our model give us
this result (see figure \ref{U1}): An $\Oplus$ -plane can be thought (at first approximation) as an $O3^-$ plus a $NS5$-brane wrapping
$\IRP^2$. Since $H_{NS}$ is non-trivial, the presence of a $\Dseven$ instantonic brane induces the creation of a $D5$-brane wrapping
$\IRP^2$. So, an $\Oplus$ transforms into $\Otplus$ via $\Dseven$, and $\tau$ goes to $\tau -1=\tau'$. Later, $\Otplus$ intersects a $(0,-1)$
seven-brane. Since $G_3$ is non-trivial a $NS5$-brane wrapping $\IRP^2$ emerges from the instantonic brane. However, since a $NS5$-brane was
already present before facing the instantonic $(0,-1)$ seven-brane, we get two $NS5$-branes giving us a zero total charge (remember that they
have a $\IZdos$ charge) and a zero $H_{NS}$-flux. Another way to say the same, is that a $NS5$-brane decays into vacuum in the presence of
the instantonic $(0,-1)$ seven-brane. So, an $\Otplus$ transforms into a $\Otminus$ and the complex coupling $\tau'$ goes now to 
$\frac{\tau'}{\tau '+1}=\tau''\;=\;\frac{\tau -1}{\tau}$. Finally, the $\Otminus$-plane ends on a $\Dseven$. Here $H_{NS}$ is trivial, and
$D5$-wrapping $\IRP^2$ does not decay into the vacuum, but $\tau"$ is transformed into $\tau"-1=\frac{-1}{\tau}$.

\begin{figure}
\begin{center}
\centering
\epsfysize=7cm
\leavevmode
\epsfbox{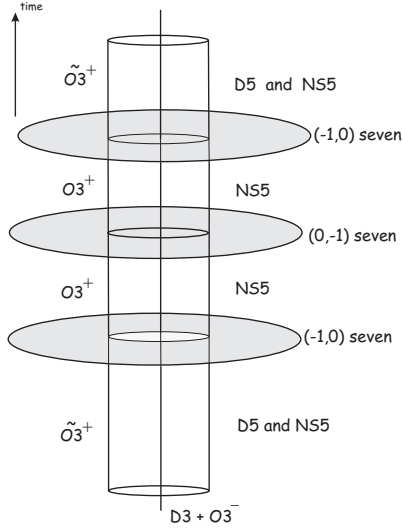}
\end{center}  
\caption[]{\small In the presence of the three seven-branes which generate $\cal S$-duality, the orientifold plane $\Otplus$ is invariant, as
expected.}
\label{U2}   
\end{figure}

Then, the final picture is that an $\Oplus$ plane is transformed into $\Otminus$ via the presence of three instantonic seven-branes, and
$\tau$ goes to $\frac{-1}{\tau}$. This is of course, S-duality on the field content coming from a $D3$-brane on top of the orientifolds. A
similar argument holds for $\Otplus$ being invariant. See figure \ref{U2}.

\begin{figure}
\begin{center}
\centering
\epsfysize=7cm
\leavevmode
\epsfbox{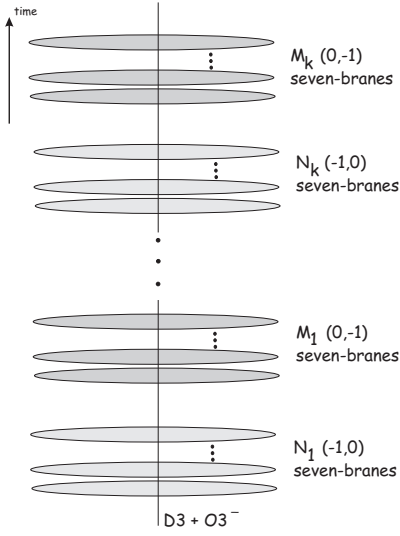}
\end{center}  
\caption[]{\small A general $SL(2,\IZ)$ transformation generated by a chain of $M_i$ $\Dseven$ and $N_i$ $(0,-1)$ seven-branes.}
\label{MN1}   
\end{figure}

Now, let us formalize the above results in terms of linear maps given by $d_3$. First of all, as it is well-known, a general transformation
under $SL(2,\IZ)$ can be expressed as a chain of transformations $\cal T$ and $\cal S$. In our case, the generators are given by ${\cal
T}^{-1}$ and $\cal U$. Notice that in our context, the transformation $\cal T$ is not a basic generator, since it implies the presence of a
$D7$-brane, that as we saw, it is inconsistent. Let us say that a general $SL(2,\IZ)$ transformation is given by the chain of generators
\beqa
{\cal U}^{M_k}\cdot {\cal T}^{-N_k}\cdot {\cal U}^{M_{k-1}}\cdot {\cal T}^{-N_{k-1}}\cdots {\cal U}^{M_0} {\cal T}^{-N_0}\;,
\eeqa
where $M_i$ and $N_i$ ($i=0,...,k$) are integers $\geq 0$. In terms of instantonic branes, $M_i$ refers to the number of $(0,-1)$ seven-branes
and $N_i$ represents the number of $\Dseven$-branes (one following the other). See picture \ref{MN1}.

We actually can compute the total effect given by the set of seven-branes by an iterating process. Let us define $\widetilde{C_0}^{M_i}$ as
the $M_i$ units of the scalar field which couples to the $(0,-1)$ seven-brane and $C^{N_i}_0$ the correspondent scalar field coupling to
$\Dseven$-branes. As we know, the presence of such instantonic branes alters the number of units of the fluxes $H_{NS}$ and $G_3$. Consider
for instance the first set of $N_0$ $\Dseven$-branes. Their effect is to give $N_0$ units of $G_3$ for each unit of $H_3$ present in the
system (or in other words, if an instantonic seven-brane is wrapping a cycle which contains a subcycle supporting $a$ units of $H_{NS}$, a
flux of $a$ $D5$-branes wrapping $\IRP^2$ must be emanating from the seven-brane). Let say we start with $a$ units of $H_3$ and $b$
units of $G_3$ (keep in mind that $a$ and $b$ are defined as integers modulo 2). The action of the first set of $\Dseven$-branes is given by
the map
\beqa
\widetilde{d_3^R} C_0^{N_0}\;=\;C_0^{N_0}\,\cup\,(a +b)\;=\;-N_0a\;,
\eeqa
from where we can read off that the effect of having non-trivial units of $G_3$ is null. The units of $H_{NS}$ and $G_3$ after the presence
of the seven brane change as: 
\beqa
(b,a) \rightarrow (b-N_0a, a)\;=\;(b+\widetilde{d_3^R}C_0^{N_0}, a)\;.
\eeqa
Similarly, the first set of $M_0$ $(0,-1)$
seven-branes, is given by the map
\beqa
\begin{array}{ccl}
\widetilde{d_3^{NS}}\widetilde{C_0}^{M_0}&=&\widetilde{C_0}^{M_0}\,\cup\,\left((a)_{H_{NS}} + (b+\widetilde{d_3^R}C^{N_0}_0)_{G_3}\right)\\
                        &=&\widetilde{C_0}^{M_0}\,\cup\, \left((a)_{H_{NS}} +(b-N_0a)_{G_3}\right)\\
			&=&\left(M_0(-N_0a+b)\right)_{H_{NS}}\;.
\end{array}
\eeqa
where $\widetilde{d_3^{NS}}\widetilde{C_0}^{M_0}$ gives us units of $H_{NS}$ and the subscripts label the units of $H_{NS}$ and $G_3$ fluxes respectively. 
Notice that the number of units of $H_{NS}$ depends of the output coming from the 
set of $N_0$ $\Dseven$-branes. The total amount of units of $G_3$ and $H_{NS}$ transforms after the presence of the seven-branes as:
\beqa
(b+\widetilde{d_3^R}C_0^{N_0}, a) \rightarrow (b+\widetilde{d_3^R}C_0^{N_0}, a+\widetilde{d_3^{NS}}\widetilde{C_0}^{M_0})\;.
\eeqa
By iterating the process, we
get that the final sets of $N_k$ $\Dseven$ and $M_k$ $(0,-1)$ seven-branes establish an output given by,
\beqa
\begin{array}{ccl}
\widetilde{d_3^R}C_0^{N_k}&=&C_0^{N_k}\,\cup\,\left((a+\widetilde{d_3^{NS}} \sum_{i=0}^{k-1}\widetilde{C_0}^{M_i})_{H_{NS}}
+(b+\widetilde{d_3^R}\sum_{i=0}^{k-1}C_0^{N_i})_{G_3}\right)\\
	    &=&-N_k\left(a+\widetilde{d_3^{NS}}\sum^{k-1}_{i=0}\widetilde{C_0}^{M_i}\right)_{G_3}\;,
\end{array}
\eeqa
where the last equality express the amount of units of $G_3$ generated by the set of $N_k$ $\Dseven$-branes. The total units of $G_3$ and
$H_{NS}$ after the action of this set of seven-branes is 
\beqa
(b+\widetilde{d_3^R}\sum^{k}_{i=0}C_0^{N_i}\;,\; a+\widetilde{d_3^{NS}}\sum^{k-1}_{i=1}\widetilde{C_0}^{M_i})
\eeqa
and
\beqa
\widetilde{d_3^{NS}}\widetilde{C_0}^{M_k}&=&\widetilde{C_0}^{M_k}\,\cup\,\left((a+\widetilde{d_3^{NS}} \sum_{i=0}^{k-1}\widetilde{C_0}^{M_i})_{H_{NS}}
+(b+\widetilde{d_3^R}\sum_{i=0}^{k}C_0^{N_i})_{G_3}\right)\\
	    &=&M_k\left(b+\widetilde{d_3^R}\sum^{k}_{i=0}C_0^{N_i}\right)_{H_{NS}}
\eeqa
where also, the last equality gives us the amount of units of $H_{NS}$-flux generated by the instantonic set of $M_k$ $(0,-1)$ seven-branes.
The total units of $(G_3, H_{NS})$ is,
\beqa
(b+\widetilde{d_3^R}\sum^{k}_{i=0}C_0^{N_i}\;,\; a+\widetilde{d_3^{NS}}\sum^{k}_{i=1}\widetilde{C_0}^{M_i})
\eeqa

\subsection{A comment on instantonic dyonic seven-branes}
Here we want to study the linear maps generated by the presence of an instantonic dyonic seven-brane. Since the full symmetry $SL(2,\IZ)$ is
generated by $\Dseven$ and $(0,-1)$ seven-branes in principle we do not require the study of dyonic seven branes, since their effects can be
expressed in terms of a set of the mentioned  seven-branes. However, it is interesting to study this case. For that we will use S-duality
since we already have obtained it from the above pictures. 

A dyonic $(p,q)$ seven-brane is gathered from a $D7$ applying a generic $SL(2,\IZ)$ transformation given by the matrix
\beqa
\left(
\begin{array}{cc}
p&r\\
q&s
\end{array}
\right)
\eeqa
where by construction, $p$ and $q$ are relative prime numbers. The monodromy $\tau$ gets by the presence of a dyonic $(p,q)$ seven-brane is
given by the matrix
\beqa
\left(
\begin{array}{cc}
-(1+pq)& p^2\\
-q^2& pq-1
\end{array}
\right)\;.
\eeqa
Let us label the scalar field which couples to this dyonic seven-brane as $C^{(p,q)}_0$. Actually, 
\beqa
C_0^{(p,q)}\,=\,\frac{(pC_0+r)(qC_0+s)+pqe^{-2\phi}}{(qC_0+s)^2+q^2e^{-2\phi}}\;.
\eeqa
Hence, if we start with $a$ units of $H_{NS}$ and $b$ units of $G_3$ fluxes, the differential linear map $\widetilde{d}_3$ acts over $C^{(p,q)}_0$ as
\beqa
\begin{array}{ccl}
\widetilde{d_3}C^{(p,q)}_0&=&C^{(p,q)}\cup (a+b)\\ 
              &=&(-q^2b+pqa)_{H_{NS}} + (-pqb+p^2a)_{G_3}\;,
\end{array}
\eeqa
and the final pair of fluxes is given by
\beqa
(b,a)\rightarrow \left((1-q^2)b+ ap^2, -q^2b +(pq-1)a\right)\;.
\eeqa
Physically what this map is telling us is that $(-(pq)b +p^2a)$ units of $D5$ and $(-q^2a +pqa)$ units of $NS5$-branes are emanating from a
dyonic $(p,q)$ seven-brane wrapping $\IRP^5$ (transverse to an orientifold three-plane)
which contains a subcycle $\IRP^3$ supporting $a$ units of $H_{NS}$ and $b$ units of $G_3$ . In order to get a picture of orientifold planes
transforming by the presence of an instantonic dyonic seven-brane, we must give values to $a$ and $b$. For instance, if $a=b=1$, the above
map tells us that an $\Otplus$-plane transforms into an $\Oplus$ plane in the presence of an $(-2,-1)$ seven-brane. Notice also that in our
case $p$ and $q$ must be negative integers since the dyonic seven-brane is a $SL(2,\IZ)$ transformation of an $\Dseven$ which can also be
expressed as an $(-1,0)$ seven-brane.

Of course, any dyonic seven-brane can be split out in a set of $\Dseven$ and $(0,-1)$ seven-branes from we can notice that a $(-p,-q)$
seven brane is not a set of $p$ $\Dseven$ and $q$ $(0,-1)$ seven-branes.

\section{$SL(2,\IZ)$ self-duality for $G=SO(2N)$, $U(N)$}
In this section we give an heuristic explanation of the self-duality in four-dimensional ${\cal N}=4$ SYM theories with gauge groups
$SO(2N)$ and $U(N)$. The former one is gathered in the low energy limit of the worldvolume field content of $D3$-branes parallel to
$\Ominus$. According to our model, the presence of instantonic seven-branes establishes a monodromy in $\tau$. However, since an
$\Ominus$-plane does not carry non-trivial discrete fluxes $G_3$ or $H_{NS}$ there are not $D5$ or $NS5$-branes created by the instantonic
seven-branes. So, in this context, an $\Ominus$-plane is invariant under the presence of such instantonic branes. This also means that the
map $\widetilde{d_3}:H^0(\IRP^5;\IZ) \rightarrow H^3(\IRP^5;\ItZ)$ is trivial, since 
\beqa
\widetilde{d_3}(C_0)=C_0\cup H_{NS}=\widetilde{d_3}\widetilde{C_0}=\widetilde{C_0}\cup G_3 =0\;.
\eeqa
Henceforth, under this pont of view, an $\Ominus$-plane is not affected by the presence of instantonic seven-branes. This means that a
seven-brane wrapped on $\IRP^5$ (and non supporting fluxes $G_3$ or/and $H_{NS}$) is an anomaly-free system and it is not necessary to add
magnetic sources to it. However, their presence do affect the complex coupling $\tau$. The monodromy in $\tau$ depends, as we saw, on
the number of $\Dseven$ and $(0,-1)$-seven-branes we have in a row. The interpretation of this issues in the field content
of $N$ $D3$-branes parallel to $\Ominus$ is that the four dimensional ${\cal N}=4$ SYM theory with gauge group $SO(2N)$ is self-dual under
the full symmetry group $SL(2,\IZ)$. Notice, that the presence of the instantonic seven-branes is once again fixed by the fact that
$H^0(\IRP^5;\IZ)$ is non-trivial.

In a similar way, let us now study the $SL(2,\IZ)$ self-duality for the gauge group $U(N)$. In this case, we do not have an orientifold
three-plane. This is an important difference since now the topology of the space transverse to $N$ $D3$-branes is $\IR^6$, and we require to
study the homology groups of this space. However, as same as we did it for the orientifolds case, we must remove the origin of
$\IR^6$, i.e., the point where $D3$-branes are placed. The resulting space is $\IR^6/\{0\}$ which can be compactified to $\IS^5$. The
non-trivial homology groups for $\IS^5$ are $H_0(\IS^5;\IZ)=H^5(\IS^5;\IZ)=\IZ$. Hence, following the same construction we showed in section
2, we want to built up three-dimensional spaces by wrapping type IIB $D$-branes on non-trivial homology cycles. The only option is to wrap
$D3$-branes on a zero-cycle (this is again an elaborated way to realize the presence of $D3$-branes), but also we can wrap five- and
seven-branes on $\IS^5$ to get lower-dimensional objects than the $D3$-branes. So, in particular we also have instantonic seven-branes
(respect to the transverse space to $D3$) which could in principle, establish a monodromy for $\tau$.

If such seven-branes are present they will  create or annihilate some other objects if there is a non-trivial map $d_3: H^p(\IS^5;\IZ)
\rightarrow H^{p+3}(\IS^5;\IZ)$ (notice also that in this case there are not twisted forms) for some $p$. But as it was said, this is not the
case. Then, although instantonic seven-branes can be present in the system, they are anomaly-free objects and no magnetic sources are
required to be added. However, the presence of these instantonic branes do establish a monodromy in $\tau$ giving us an $SL(2,\IZ)$
transformation.

Summarizing, the topology of the space transverse to $D3$-branes, allows the presence of an instantonic seven-brane, but there are not
objects which decays into vacuum or that are created from vacuum by their presence. This implies in the point of view of the field content
from the three-branes, that the ${\cal N}=4$ SYM theory with gauge group $U(N)$ is invariant under $SL(2,\IZ)$ transformations. In this case
we conclude that $SL(2,\IZ)$ duality is the physical interpretation of the fact that cohomology is indeed the final step in the approximation
to a K-theory classification.

A note to be added: In the present case, it is also possible to have instantonic five branes (Dirichlet or Neveu-Schwarz) supporting a
non-trivial amount of fluxes (RR or NS-NS) which are not related to the presence of orientifolds (they are integer units). This case has
been studied in detail in \cite{JE3, JE1, JE2}.

\section{Conclusion}
In this paper we have applied the Maldacena-Moore-Seiberg (MMS) physical interpretation of the Atiyah-Hirzebruch Spectral Sequence (AHSS) in
a background with orientifold three-planes. We also extended the AHSS to include a classification of NS-NS fluxes. 

In general, cohomology tells us that there are four different types of orientifold three-planes: $\Oplus, \Otplus, \Ominus$ and $\Otminus$.
In the case of a classification of RR fields, the AHSS shows that $\Oplus$ and $\Otplus$ are equivalent in K-theory. This equivalence is
given by the surjective map
\beqa
\widetilde{d_3}:H^0(\IRP^5;\IZ)=\IZ\rightarrow H^3(\IRP^5;\ItZ)=\IZdos\;,\nonumber
\eeqa
where the two classes in $H^3$ (classifying $\Oplus$ and $\Otplus$) are exact and lifted to a trivial class in K-theory, as was pointed out 
in \cite{BGS}. According to the picture given by MMS, the equivalence is explained as follows: a $D5$-brane wrapping $\IRP^5$ (where the
cycles are transverse to an $\Oplus$-plane) ends on an instantonic anti-seven-brane wrapping $\IRP^5$. Since the cohomology version of an
$\Otplus$ is given by a $D5$-brane wrapped on $\IRP^5$ transverse to $\Oplus$, this picture tells us that an $\Otplus$ is transformed into an
$\Oplus$ via the instantonic brane.

The consequence on the field theory gathered from $D3$-branes on top of $\Oplus$ is that the complex coupling $\tau$ acquires a non-trivial
monodromy, $\tau \rightarrow \tau -1 $,  in the presence of the seven-brane. Hence the equivalence among orientifolds is translated into the ${\cal T}\in SL(2,\IZ)$ duality
of ${\cal N}=4$ SYM.

We also extended the map $\widetilde{d_3}$ to classify NS-NS fluxes as well. This was done by using the fact that both fields $G_3$ and
$H_{NS}$ are classified by $H^3(\IRP^5,\IZ)$. We found that $\Otplus$ and $\Otminus$ must be equivalent in a twisted K-theory
classification (of which we do not say anything). The physical picture of this equivalence involves an instantonic $(0,-1)$ seven-brane
($\cal S$-dual of a anti $D7$) wrapping $\IRP^5$, transverse to an $\Otminus$, where a $NS5$-brane wrapping $\IRP^2$ ends. Since an
$\Otminus$ plus a $NS5$-brane wrapping $\IRP^2$ is the cohomology version of $\Otplus$, we conclude that an $\Otplus$ transforms into an
$\Otminus$ in the presence of the instantonic seven-brane.

Once again, the consequence on the field theory on $D3$-branes on top of the orientifolds is non-trivial. This is due to the fact that
first, the gauge groups associated to the above two orientifolds are different ($SO(2N+1)$ and $USp(N)$), and second, that $\tau$ has also a
non-trivial monodromy in the presence of an $(0,-1)$ seven-brane.
We found that the above equivalence is interpreted as the ${\cal U}\in SL(2,\IZ)$ duality ($\tau \rightarrow \tau/(\tau+1)$) for ${\cal N}=4$
SYM theories.

Hence, with these two equivalences we were able to generate the full $SL(2,\IZ)$ symmetry in the field content, as a consequence of the
non-triviality of the mapping $\widetilde{d_3}$. Summarizing, we conclude that:
\begin{enumerate}
\item
By extending the AHSS to include NS-NS fluxes as well as RR ones, there are two `basic' equivalences among orientifold three-planes. One is
given throughout  a classification of RR fields, which gives us the equivalence between $\Oplus$ and $\Otplus$, and the second is gathered
from a classification of NS-NS fluxes, giving us the result that $\Otplus$ and $\Otminus$ are equivalent.
\item
Combining the above two equivalences, we get that the four $O3$-planes classified by cohomology are lifted to only two planes in K-theory.
This means that three of them are equivalent to each other. However, since ${\cal N}=4$ SYM theories are gathered from $D3$-branes on top of
these four planes, such equivalences must imply a relation between them. This relation is indeed the $SL(2,\IZ)$-duality of ${\cal N}=4$ SYM.
Also we are able to explain that in the case of the gauge groups $U(N)$ or $SO(2N)$, the $SL(2,\IZ)$ duality of the field theory follows from
the fact that $\Ominus$  and a $D3$-brane (in a background without orientifold planes), are both non equivalent to other different object. A
similar result was found in reference \cite{gerbes}, where the author classified orientifolds (in particular orientifold-three planes) by
gerbes.
\item
Our main conclusion (at least for the present case) is that $SL(2,\IZ)$ duality of ${\cal N}=4$ SYM is a consequence of the topology of the
space. When two different cohomology classes belonging to the same group are lifted to a trivial class in K-theory, two different objects are
said to be equivalent in K-theory. This equivalence is interpreted in the field content (on $D$-branes) as a duality among the two
systems. Such a duality is performed throughout the presence of an instantonic brane. When this brane does not  relate two different systems,
we say that the correspondent field theory is self-dual under such a symmetry.
\end{enumerate}

\begin{center}
{\bf Acknowledgments}
\end{center}
I would like to thank J. Evslin, A. Hanany, C. Nu\~nez, D. Tong, J. Troost and specially to Alfredo Iorio for very useful suggestions and discussions, and the
CTP for kind hospitality.
This work is
supported in part by a CONACyT (M\'exico) fellowship with number 020136 and by the U.S. Department of Energy (D.O.E.) under cooperative
research agreement \#DF-FC02-94ER40818. \\

\appendix
\section{The system $Op$-$D(p+6)$}
In this section we give a briefly review on the required conditions to have an stable system performed by an $O3$-plane in coordinates 0123, 
and a $D7$-brane extending along coordinates 01345678. This
system has six Dirichlet-Neumann directions. So, let us start by analyzing a system conformed by a
$Dp$ and a $Dp'$ -brane. 

The cylinder amplitude $\cal A$ is given by
\beqa
{\cal A}\,\sim\,Tr\;\frac{1}{2}(1+(-1)^F)e^{-2\pi\alpha t}\;,
\eeqa
where $H$ is the Hamiltonian of the system. In terms of the following functions,
\beqa
\begin{array}{ccl}
f_1(q)&:=& q^{\frac{1}{12}}\prod^\infty_{n=1}(1-q^{2n})\\
f_2(q)&:=& 2^{1/2}q^{1/12} \prod^{\infty}_{n=0}(1+q^{2n})\\
f_3(q)&:=& q^{-1/24}\prod^\infty_{n=1}(1+q^{2n-1})\\
f_4(q)&:=& q^{-1/24}\prod^\infty_{n=1}(1-q^{2n-1})
\end{array}
\eeqa
with $q=e^{-\pi t}$, we get that the amplitude is
\beqa
{\cal A}\sim \left(\frac{1}{2}[f^{8-\nu}_3(q)
-f_4^{8-\nu}(q)][2^{\nu/2}f_2^\nu(q)]\right)_{NS-NS}-\left(\frac{1}{2}2^{(8-\nu)/2}f_2^{8-\nu}(q)[f^\nu_3(q)-f_4^\nu(q)]\right)_{RR}\;,
\eeqa
where $\nu=|p-p'|$ and the subscripts refer to the contribution of NS-NS and RR fields to the force among these two $D$-branes, in the close
string channel. It is possible to show (see \cite{polchinski}) that for $\nu = 6$ the RR force is stronger than the NS-NS one and as a
consequence, the system is repulsive
at short and long distances. In order to get an stable system at any distance, the RR contribution to the
amplitude must be weaker than the NS-NS one. This can be given by a system formed by a $\Dseven$ and a $D3$-brane. 

In a similar way, a $D7$ and an $O3$-plane will be unstable (repulsive). 
All different possibilities are given in table \ref{D7O3}, where a minus (plus) sign refers to a repulsive (attractive) force.
\begin{table}
\begin{center}
\caption{Sign of the amplitude between seven-branes and $O3$-planes.} 
\label{D7O3}
\begin{tabular}{||c|c||c|c||c||}\hline\hline
& & RR & NS-NS & Total\\\hline\hline
$D7$&$\Ominus$&+&$-$&+\\
$D7$&$\Oplus$&$-$&+&$-$\\
$\Dseven$&$\Ominus$&$-$&$-$&$-$\\
$\Dseven$&$\Oplus$&+&+&+\\\hline\hline
\end{tabular}
\end{center}
\end{table}
However, if we add $N$ units of $D3$-branes on top of the orientifold three planes, all systems conformed by $O3$ and $N$ $D3$-branes will
have a positive RR charge (including an $\Ominus$-plane). Table \ref{D7O32} shows the sign of RR force between different objects.
\begin{table}
\begin{center}
\caption{Sign of the amplitude between seven-branes and $O3$-planes with $N$ $D3$-branes.} 
\label{D7O32}
\begin{tabular}{||c|c||c||}\hline\hline
& & RR \\\hline\hline
$D7$&$\Ominus + N\,D3$&$-$\\
$D7$&$\Oplus$, $\Otplus$, $\Otminus$ + $N$ $D3$&$-$\\
$\Dseven$&$\Ominus + N\, D3$&+\\
$\Dseven$&$\Oplus$, $\Otplus$, $\Otminus$ + $N$ $D3$&+\\\hline\hline
\end{tabular}
\end{center}
\end{table}
Then, we can see that in order to have stable systems (attractive), we must consider an anti-seven brane $\Dseven$ instead of a positive RR
charge one ($D7$). This is the reason we work with an $\Dseven$ as the instantonic brane present in our models.

\section{The AHSS}
In this section we give a briefly review about the AHSS.
The AHSS is an algebraic algorithm that relates K-theory to integral cohomology (see \cite{DMW} and references therein). The basic idea is to compute $K(X)$ by using a
sequence of successive approximations. First, a filtration on $K(X)$ is introduced by defining $K_p$ as the kernel of $K(X)$ on $K(X^p)$,
where $X^p$ is the $p$-skeleton of $X$. The spectral sequence computes the graded complex $GrK(X)$ given by
\beqa
GrK(X)=\oplus_pE^p_r = \oplus_p K_p(X)/K_{p+1}(X)\;,
\eeqa
where $r$ stands for the order of the approximation. In each step of approximation, classes that are not closed under the map $d^p_r: E^p_r
\rightarrow E^{p+r}_r$ are removed. On the other hand, classes that are closed survive the refinement while exact classes are mapped to the
trivial class in the next step. The first non-trivial approximation is given by integer cohomology groups,
\beqa
E= \oplus_p E^p_r =\oplus_p H^p(X,\IZ)\;,
\eeqa
where $p$ is odd (even) for type IIB (IIA) string theory. Hence, each step of approximation $E^p_{r+1}$ is given by
\beqa
E^p_{r+1}=ker\:d_r/Im\:d^{p-r}_r\:.
\eeqa
Roughly speaking, we can say that K-theory classes are given by the above quotient. In the case of string theory, the only non-trivial maps
of higher order than one, are given by $d_3$ and $d_5$, although the latter is trivial in most of the cases \cite{BGS, M, JE3, JE1, JE2}. The mapping $d_3$ is
given explicity by 
\beqa
d_3 = Sq^3 +[H_{NS}]\:,
\eeqa
where $Sq^3$ is the Steenrod square and $[H_{NS}]$ is the cohomology class of the NS-NS field strength $H_{NS}=dB_{NS}$. Actually, the Steenrod square
is also related to the third Stiefel-Whitney class (or the Bockstein map of the second Stiefel-Whitney class) denoted as $W_3({\cal W})$,
with $\cal W$ being a submanifold of $X$. When $W_3({\cal W}) \in \IZdos$ is the zero class, we say that $\cal W$ is an spin$^c$ manifold.

For real K-theory (or in general, for K-theory groups with freely acting involutions) the
approximations are given by {\it twisted} or {\it untwisted} maps (see appendix in \cite{BGS}), i.e., the differential operator 
$\widetilde{d_3}$
maps twisted into untwisted classes and vice versa. In this case
$\widetilde{d_3}=\widetilde{Sq^3}+H_{NS},$ with $[H_{NS}]\in \IZdos$. It is found that $\widetilde{Sq^3}$ is
trivial for both values of $\IZdos$ and $d_5$ is trivial in all cases.
$d_5$ maps (un)twisted into (un)twisted classes.

The first approximation to the graded complex $GrK^{-s}(X)$$=\bigoplus_n E^{p,-(p+s)}_n$,
with

\beqa
E^{p,-(p+s)}_n(X)\,=\,K^{-s}_p(X)/K^{-s}_{p+1}(X),
\eeqa
is given by
\beqa
\begin{array}{ccl}
E^{p,q}_1&=&C^p(X|_{\tau},\IZ)\; \hbox{for $q = 0\;{\rm mod}\;4$}\\
E^{p,q}_1&=&C^p(X|_{\tau},\ItZ)\; \hbox{for $q = 2\;{\rm mod}\;4$}\\
E^{p,q}_1&=&0\;\hbox{for $q$ odd,}
\end{array}
\label{ahss0}
\eeqa
where $\tau$ is the freely acting involution on $X$. Then, the second order of this approximation is 
given by 
the cohomology groups

\beqa
\begin{array}{ccl}
E^{p,q}_2&=&H^p(X|_{\tau},\IZ)\; \hbox{for $q = 0\;{\rm mod}\;4$}\\
E^{p,q}_2&=&H^p(X|_{\tau},\ItZ)\; \hbox{for $q = 2\;{\rm mod}\;4$}\\
E^{p,q}_2&=&0\;\hbox{for $q$ odd}.
\end{array}
\label{ahss}
\eeqa
These results are valid for orthogonal and quaternionic K-theory groups.



\bigskip

\begin{thebibliography}{99}


\bibitem{WB}
E. Witten, ``Baryons And Branes In anti de Sitter Space'', JHEP 9807
(1998) 006, hep-th/9805112.

\bibitem{HO}
A. Hanany and B. Kol, ``On Orientifolds, Discrete Torsion, Branes and M Theory'',
JHEP 0006 (2000) 013, hep-th/0003025.

\bibitem{HT}
A. Hanany and J. Troost, ``Orientifold Planes, Affine Algebras and Magnetic Monopoles", JHEP 0108:021,2001, hep-th/0107153.

\bibitem{BGS}
O. Bergman, E. Gimon and S. Sugimoto, ``Orientifolds, RR Torsion, and K-theory'',
JHEP 0105 (2001) 047, hep-th/0103183.

\bibitem{M}
J. M. Maldacena, G.W. Moore and N. Seiberg, ``D-Brane Instantons and K theory Charges", JHEP 0111:062,2001, hep-th/0108100.

\bibitem{FW}
D.S. Freed and E. Witten, ``Anomalies in String Theory with D-branes'',
hep-th/9907189.

\bibitem{Wk}
E. Witten, ``D-branes and K-theory'', JHEP 9812 (1998) 019, hep-th/9810188.

\bibitem{DMW}
D-E. Diaconescu, G.W. Moore and E. Witten, ``$E(8)$ Gauge Theory, and a Derivation of K Theory From M Theory", 
Adv.Theor.Math.Phys.6:1031-1134,2003, hep-th/0005090.

\bibitem{GK}
A. Giveon and D. Kutasov, ``Brane Dynamics and Gauge Theory", Rev.Mod.Phys.71:983-1084,1999, hep-th/9802067.

\bibitem{Be}
X. Bekaert, ``Issues in electric-magnetic duality", PhD Thesis,  hep-th/0209169.

\bibitem{USP32}
S. Sugimoto, ``Anomaly cancellations in Type I D9-anti-D9 system and the $USp(32)$ string theory'', 
Prog.Theor.Phys.102:685-699,1999, hep-th/9905159.

\bibitem{braneboxes} A. Hanany and E. Witten,``Type IIB Superstrings, BPS Monopoles,
and Three-dimensional Gauge Dynamics'', Nucl. Phys. B 492 (1997) 152, hep-th/9611230; \\
A. Hanany and A. Zaffaroni, ``On the Realization of Chiral Four-dimensional Gauge
Theories Using Branes', JHEP 9805 (1998) 001, hep-th/9801134.

\bibitem{HIS}
Y. Hyakutake, Y. Imamura and S. Sugimoto, ``Orientifold Planes, Type I Wilson Lines
and Non-BPS D-branes'', JHEP 0008 (2000) 043, hep-th/0007012.

\bibitem{yo}
O. Loaiza-Brito and A.M. Uranga, ``The Fate of the Type I Non-BPS D7 Brane'', Nucl.
Phys. B 619 (2001) 211, hep-th/0104173.

\bibitem{yo2}
H. Garc\'{\i}a-Compe\'an and O. Loaiza-Brito, ``Branes and Fluxes in Orientifolds and K-theory", hep-th/0206183.


\bibitem{monopoles}
J. A. Harvey, ``Magnetic Monopoles, Duality, and Supersymmetry", Prepared for ICTP Summer School in High-energy Physics and Cosmology, 
Trieste, Italy, 12 Jun - 28 Jul 1995.
Published in Trieste HEP Cosmology 1995:66-125 (QCD161:W626:1995) Also in *Boulder 1996, Fields, strings and duality* 157-216. 
hep-th/9603086.\\
D. Olive, ``Exact Electromagnetic Duality", Nucl.Phys.Proc.Suppl.45A:88-102,1996, Nucl.Phys.Proc.Suppl.46:1-15,1996, hep-th/9508089.\\
P. Di Vecchia, ``Duality in supersymmetric gauge theories",  Surveys High Energ.Phys.10:119-151,1997, hep-th/9608090.\\
S. V. Ketov, ``Solitons, Monopoles and Duality: from Sine-Grodon to Seiberg-Witten", Fortsch.Phys.45:237-292,1997, hep-th/9611209.

\bibitem{EGKT}
S. Elitzur, A. Giveon, D. Kutasov and D. Tsabar, ``Branes, Orientifolds and Chiral Gauge Theories". Nucl.Phys.B524:251-268,1998, 
hep-th/9801020.

\bibitem{U}
A.M. Uranga, ``Comments on Nonsupersymmetric Orientifolds at Strong Coupling'', JHEP
0002 (2000) 041, hep-th/9912145. 

\bibitem{Mo}
G. Moore, ``K-theory from a Physical Persepctive", hep-th/0304018.

\bibitem{JE3}
J. Evslin, ``Twisted K-theory from Monodromies", JHEP 0305:030,2003, hep-th/0302081.

\bibitem{urangak}
A. M. Uranga, ``D-brane Probes, RR Tadpole Cancellation and K Theory Charge", Nucl.Phys.B598:225-246,2001 ,hep-th/0011048.

\bibitem{JE1}
J. Evslin and U. Varadarajan, ``K-theory and S-duality:Starting Over from Square 3", JHEP 0303:026,2003, hep-th/0112084.

\bibitem{JE2}
J. Evslin, ``IIB Soliton Spectra with All Fluxes Activated", Nucl.Phys.B657:139-168,2003, hep-th/0211172.

\bibitem{sugraeqns}
M. B. Green, C. M. Hull and P. K. Townsend, ``D-Brane Wess-Zumino Actions, T-duality and the Cosmological Constant", 
Phys.Lett.B382:65-72,1996, hep-th/9604119.

\bibitem{urangac}
J. F. Cascales and A. Uranga, ``Chiral 4D string vacua with D-branes and NSNS and RR fluxes", JHEP 0305:011,2003, hep-th/0303024.

\bibitem{sevenb1}
C. Vafa, ``Evidence from F-theory", Nucl.Phys.B469:403-418,1996, hep-th/9602022.

\bibitem{sevenb2}
M. Douglas and M. Li, ``D-brane Realization of ${\cal N}=2$ Super Yang-Mills Theory in Four Dimensions", hep-th/9604041

\bibitem{sevenb3}
A. Sen, ``BPS States on a Three Brane Probe", Phys.Rev.D55:2501-2503,1997, hep-th/9608005.

\bibitem{polchinski}
J. Polchinski, ``String Theory", Vol. 2; Cambridge Univ. Pr. (1998).

\bibitem{gerbes}
Arjan Keurentjes, ``Classifying orientifolds by flat n gerbes", JHEP 0107:010,2001, hep-th/0106267,\\
``Flat Connections from Flat Gerbes",
Contributed to Workshop on the Quantum Structure of Spacetime and the Geometric Nature of Fundamental 
Interactions, Corfu, Greece, 13-20 Sep 2001, hep-th/0201072.





\end{thebibliography}
\end{document}